\documentclass[twocolumn,showpacs,superscriptaddress,amsmath,amssymb,aps,prb]{revtex4}
\usepackage{times}
\usepackage{graphicx}% Include figure files
\usepackage{subfigure}
\usepackage{dcolumn}% Align table columns on decimal point
\usepackage{mathrsfs}% Math symbols
\usepackage{amsmath,bm,amsfonts}
\usepackage{latexsym}
\usepackage{textcomp}
\usepackage{pifont}
\usepackage{gensymb}
\usepackage{epstopdf}
\usepackage[perpage]{footmisc} %  for customizing footnotes
\usepackage[breaklinks,colorlinks = true,linkcolor = blue,urlcolor  = blue,citecolor = red,anchor color = green,bookmarks=true]{hyperref}

\def \be {\begin{equation}}
\def \being {\begin{equation*}}
\def \en {\end{equation}}
\def \ening {\end{equation*}}

\def \d {^\dagger}
\def \tp {t^\prime}

\begin{document}

%\title{\bf{IHM in square lattice with NNN hopping}}}
\title {\bf Correlation driven metallic and half-metallic phases in a band insulator}
\author{Soumen Bag$^{1}$, Arti Garg$^{2}$, and H. R. Krishnamurthy$^{1}$}
\affiliation{$^{1}$ Department of Physics, Indian Institute of Science, Bangalore 560 012, India \\
$^{2}$Condensed Matter Physics Division, Saha Institute of Nuclear Physics, 1/AF Bidhannagar, Kolkata 700 064, India}

\date{\today}

\begin{abstract}

We demonstrate, using dynamical mean-field theory with the hybridization expansion continuous time quantum montecarlo impurity solver, a rich phase diagram with {\em correlation driven metallic and half-metallic phases} in a simple model of a correlated band insulator, namely, the half-filled ionic Hubbard model (IHM) with first {\em and} second neighbor hopping ($t$ and $\tp$), an on-site repulsion $U$, and a staggered potential $\Delta$.  Without $\tp$ the IHM has a direct transition from a paramagnetic band insulator (BI) to an antiferromagnetic Mott insulator (AFI) phase as $U$ increases. For weak to intermediate correlations, $\tp$ frustrates the AF order, leading to a paramagnetic metal (PM) phase, a ferrimagnetic metal (FM) phase and an anti-ferromagnetic half-metal (AFHM) phase in which electrons with one spin orientation, say up-spin, have gapless excitations while the down-spin electrons are gapped. For $\tp$ less than a threshold $ t_1$,  there is a direct, first-order,  BI to AFI transition as $U$ increases, as for $\tp=0$; for $t_4< \tp < \Delta/2$, the  BI to AFI transition occurs via an intervening PM phase. For $\tp > \Delta/2$, there is no BI phase, and the system has a PM to AFI transition as $U$ increases. In an intermediate-range  $t_2 < \tp < t_3$, as $U$ increases the system undergoes four transitions, in the sequence BI $\rightarrow$ PM $\rightarrow$ FM $\rightarrow$ AFHM $\rightarrow$ AFI; the FM phase is absent in the ranges of $\tp$ on either side, implying three transitions. The BI-PM, FM-AFHM and AFHM-AFI transitions, and a part of the PM-FM transition are continuous, while the rest of the transitions are first order in nature. The PM, FM and the AFHM phases have, respectively, spin symmetric, partially polarized and fully polarized electron [hole] pockets around the ($\pm\pi/2$, $\pm\pi/2$) [($\pm \pi, 0$), ($0. \pm \pi$)] points in the Brillouin zone. 
\end{abstract}
\maketitle
\section{Introduction}

%In this paper, we study the \textit{Ionic Hubbard model} (IHM) which is an interesting extension of the Hubbard model and is perhaps the simplest model for a correlated band insulator which has been of a fair amount of interest to the condensed matter community in the last two decades or so.
\textit{The ionic Hubbard model} (IHM) is an interesting extension of the Hubbard model and is perhaps the simplest model for a correlated band insulator which has been of a fair amount of interest to the condensed matter community in the last two decades or so.
The model was first proposed for describing the neutral-ionic transition in charge-transfer organic chains\cite{nutralionic1, nutralionic7, nutralionic5, nutralionic10}. The IHM has also been invoked for understanding polarization phenomena in perovskite materials\cite{ polarization1, polarization2, polarization3, polarization4, polarization5} and covalent insulators such as FeSi and $FeSb_2$\cite{fesi}. Recently this model has also been simulated in ultra-cold atom experiments\cite{UCA1}.

Specifically, the IHM is the Hubbard model on a bipartite lattice with a staggered potential ($\Delta$) added. The staggered potential $\Delta$ doubles the unit-cell and, in the absence of $U$,  gives rise to a gap in the single particle excitation spectrum such that at half-filling the system is a band insulator (BI).  But when the Hubbard interaction $U$ is introduced, it opposes the larger occupancy of particles on one sublattice and holes on the other sublattice (favoured by $\Delta$) and prefers single occupancy on each site. Thus there is a competition between the staggered potential and the Hubbard interaction and one would expect interesting phase transitions as a function of $U/\Delta$. In particular, at half-filling, the ground state of the IHM is a band insulator (BI) for $U << \Delta$ and a Mott insulator (MI) for $U >> \Delta$.
In a Phys. Rev. Lett.~\cite{ArtiPara} co-authored by two of us, we showed using single site dynamical mean field theory (DMFT) that in the simplest IHM, with only nearest neighbour hopping $t$, a sufficiently large Hubbard $U$ can close the gap in the single particle spectrum all the way to zero, resulting in an intervening \textit{correlation induced} metallic phase provided the system is constrained to be paramagnetic. However, if anti-ferromagnetic (AF) order is allowed for, a direct transition from the BI phase to an anti-ferromagnetic insulator (AFI) phase occurs, preempting the transition to the correlation induced paramagnetic metallic (PM) phase \cite{Byczuk,ArtiHM,soumen,2D4,2D3}.   \\

There have been fairly extensive studies of the physics of intermediate values of $U$ (i.e., of order $\Delta$) in the context of the IHM with only nearest neighbour hopping  {\em in one dimension}~\cite{one6,one16,one17,one7,one8,one9,one20} which suggest that a bond-ordered (BO) insulating phase (and not a PM phase) separates the BI and MI phases.
In two and higher dimensions the existence and nature of the intermediate phase is still controversial despite several attempts based on various approximate methods of solving the (nearest neighbour hopping) IHM~\cite{2D1, 2D2, 2D3, 2D4,Graphene}. Determinantal Quantum Monte Carlo~\cite{2D1} studies suggest that the intermediate phase at half-filling is a paramagnetic metal, whereas an insulating bond ordered phase is observed in cluster DMFT (+lanczos)~\cite{2D3} studies. The variational cluster approach to the IHM at half-filling~\cite{2D4} leads to the same result as obtained by  the single site DMFT~\cite{Byczuk,soumen} approach allowing for AF order mentioned earlier,  that the system undergoes a direct transition from a  BI to an AF Mott insulator [except for a sliver of AF half-metal phase seen inside the AFI phase close to the transition point~\cite{soumen}]; however, when magnetic ordering is suppressed, single site DMFT~\cite{ArtiPara,Byczuk} shows an intermediate metallic state. Note that there is in principle a metallic point at the critical U corresponding to the transition from the BI to the BO phase even in the 1D IHM. An interesting ferrimagnetic half-metallic phase has been predicted in the IHM doped away from half-filling~\cite{ArtiHM}. A half metallic phase has been suggested in the IHM on a bilayer honeycomb lattice\cite{GrapheneHM}.

It is an obvious possibility that the introduction of frustration might inhibit the formation of AF order and stabilize the correlation induced PM phase. However, an explicit calculation which allows for AF order and nevertheless shows a stable PM phase in such a context has been missing. In this work, we  present and discuss the results of such a calculation for the IHM on a square lattice with  the inclusion of a second neighbour hopping which can frustrate the AF order. We demonstrate that for $t'/t$ sufficiently large, the system does indeed support a stable correlation induced PM phase for a range of $U$. Furthermore, as a bonus, other interesting phases now appear for ranges of $U$ that depend on the value of $t'/t$, such as a ferrimagnetic metal (FM) phase [which has non-zero values of the uniform as well as staggered magnetization],
and an anti-ferromagnetic half-metal (AFHM) phase [in which, apart from the AF order, the single particle density of states (DOS) of one of the spins is gapped where as that of the other spin is gap-less at the   Fermi level], before the system eventually turns into an AF insulator as the strength of $U$ is increased.
%\cite{ctqmcwerner, ctqmcRMP, ctqmchaule}
 The calculations on the IHM in the presence of second neighbour hopping $t^\prime$ which we discuss in this paper have been carried out mainly using single site DMFT\cite{dmftRMP} combined with the hybridization expansion continuous time quantum monte carlo (HYB-CTQMC) method\cite{ctqmc} as the impurity solver. As mentioned earlier, the goal is to get the strongly correlated stable paramagnetic metallic phase to become stable by frustrating the AFM order. The DMFT calculations have been supplemented by the much simpler (unrestricted) Hartree-Fock (HF) calculations, whose results we discuss as well. Effects arising from the inclusion of a second neighbour hopping on AF order have been studied extensively earlier in the context of the Hubbard model on various lattices~\cite{Vollhardt,duffy}, but to the best of our knowledge, its effects  on the IHM have not been explored much except for one recent work on the one-dimensional IHM\cite{1dtt}.

\begin{figure}
 \includegraphics[width=9cm]{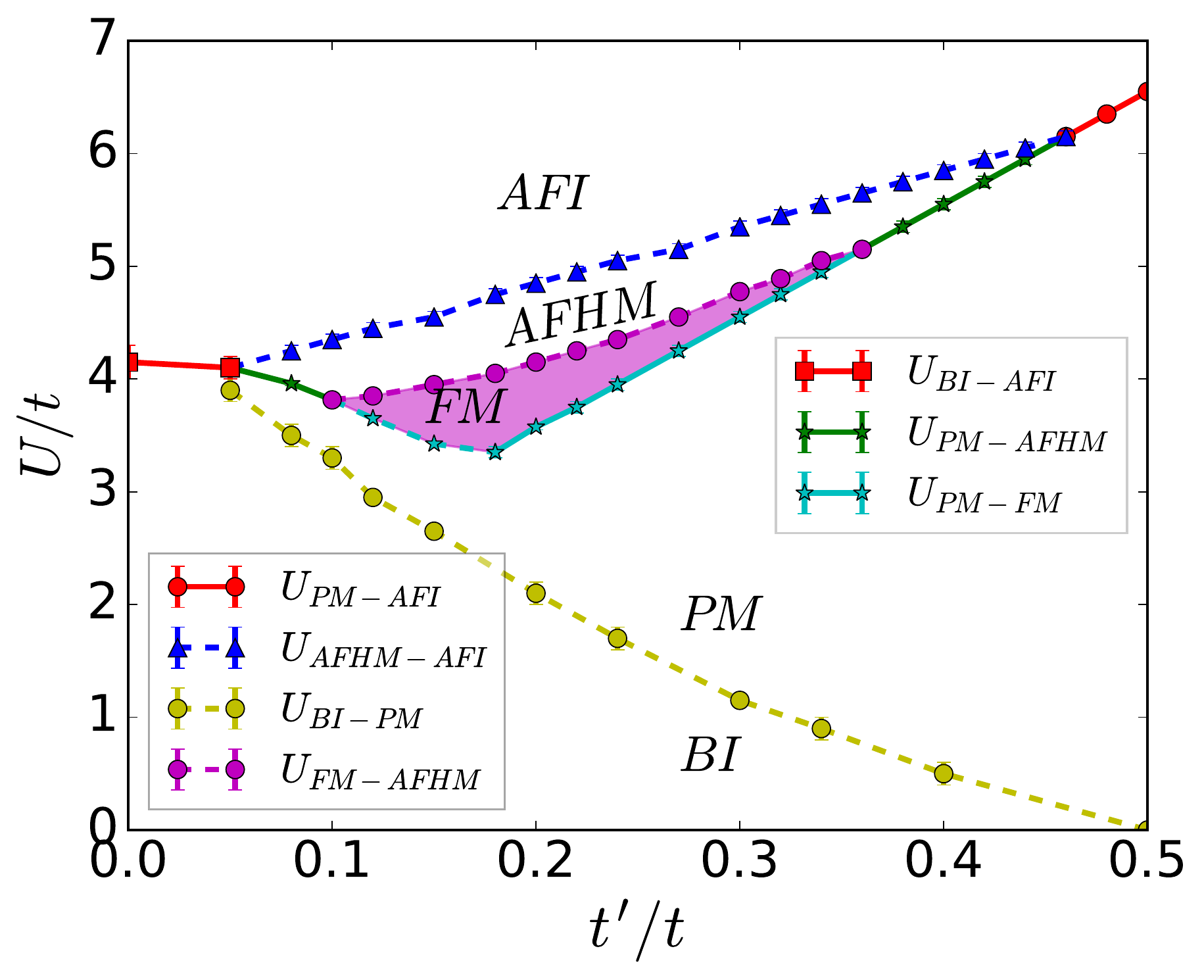}
  \caption{Phase diagram of the $t-\tp$ IHM in $\tp-U$ plane at half filling obtained using DMFT combined with CTQMC impurity solver at $\beta t$=50.0, $\Delta=1.0t$. Here BI $\equiv$ Band Insulator, PM $\equiv$ Para magnetic Metal, FM$\equiv$Ferrimagnetic Metal, AFHM$\equiv$Anti-Ferromagnetic Half Metal, AFI$\equiv$Anti-Ferromagnetic Insulator. %The line joining the points in the phase diagram labelled with the format $U_{phase1-phase2}$ means that the system goes from "phase1" to "phase2" upon increasing $U$ across the line. 
For $0.0t<\tp<0.05t$, the system goes from BI to AFI phase with increasing $U$. For $\tp>0.5t$ there is no BI phase, and the system undergoes a transtion from PM to AFI. For intermediate values of $\tp$  the system exhibits interesting half metallic phases for a range of $U$ values.  For $0.05t <\tp<0.1t$ and for $0.36 t < t'< 0.46 t$ the system shows three transitions as it goes from BI to PM to AFHM to AFI phase with increasing $U$. Whereas for $0.12t<\tp<0.36t$, the FM phase intervenes between the PM and the AFHM phases. For $0.46t<\tp<0.5t$ the system goes from BI to  PM to AFI phase with increasing $U$. 
Full lines in the phase diagram indicate first order transition lines, whereas dashed lines indicate continuous transitions.} 
\label{phase_diag}
\end{figure}

The main findings of our study are summarized in the phase diagram in Fig.~\ref{phase_diag}, drawn for $\Delta = 1.0 t$ . For $0.0<t^\prime<0.05t$, the system does indeed undergo a direct, first order transition from the paramagnetic BI to the Anti-Ferro (AF) insulator (AFI) as $U$ is increased. However, for  $0.05t<t^\prime<0.5t$, the system undergoes a transition from the paramagnetic BI to the paramagnetic metal (PM) as $U$ is increased from zero. On increasing $U$ further,  magnetic order turns on, but the nature of the  phases and phase transitions encountered differ for different ranges of  $t^\prime/t$ . For $0.05 t < t^\prime < 0.1 t $,  AF order together with a single particle excitation gap in one spin channel turn on abruptly at a threshold $U$, resulting in a first order transition from the PM phase to an AF half-metal (AFHM).  On further increasing $U$ the gapless spin channels develops a gap as well, and the system undergoes a continuous transition from AFHM to an AF (Mott) insulator (AFI). For  $0.1t<t^\prime<0.36t$, the magnetic order that turns on is initially a ferrimagnetic order , with both staggered and uniform magnetization turning on simultaneously; hence the PM phase undergoes a transition into a ferrimagnetic metal (FM) phase. As depicted in ~\ref{phase_diag}, we find that this transition is continuous for $0.1 t < t^\prime < 0.18 t$, and first order thereafter. As $U$ increases further, the uniform magnetization decreases, though the staggered magnetization keeps increasing. Coincident with the decay of the uniform magnetization to zero, a single particle excitation gap opens up in one spin channel, resulting in  a continuous transition into an AF half-metal (AFHM). On further increasing $U$ the gapless spin channel develops a gap as well, and the system undergoes a continuous transition from the AFHM into the AFI phase as before. For larger values of $t^\prime$,  $0.36t<t^\prime<0.46t$, the ferrimagnetic metallic phase disappears and the paramagnetic metal phase directly undergoes a first order transition into AFHM phase, followed by a continuous transition to the AFI phase. For even larger values of $t^\prime$, there is no AFHM phase, and the system undergoes only two transitions; first from BI to the paramagnetic metal and then from the metal to the AFI. Even the BI phase ceases to exist for $t'>0.5t$ . Amazingly, as we show in the following section, all the same phases show up in a simple unrestricted Hartree-Fock (HF) study of the $t-t'$ IHM, although the parameter values where the transitions occur , as well as some aspects of the topology of the phase diagram differ between the HF and the DMFT+CTQMC calculations. Needless to say, we expect the latter to be more accurate.

The rest of this paper is organized as follows. In Section II we present the model and its physics within a simple unrestricted HF approximation.  In Section III we present and discuss the phase diagram  and different observables calculated using DMFT+CTQMC. These are the central new results of our paper. In Section IV, we discuss (the mainly quantitative) differences in the phase diagrams obtained by the two methods. We end this paper with a concluding discussion in section V.  Some details of the calculation methods used are presented in Appendix A for the HF approximation, and in Appendix B for DMFT+CTQMC.\\

\section{IHM Hamiltonian, Simple limits and HF mean Field Theory}
In this section we present the Hamiltonian for the IHM, its physics in simple limits, and possible phases within the simplest unrestricted HF mean field theory. The Hamiltonian for the IHM can be written as
\begin{equation}
 \begin{split}
&H_{IHM} = -t\sum_{\langle ij\rangle,\sigma}(\hat{c}_{j \sigma}\d \hat{c}_{i \sigma}+ H.c) - \tp \sum_{\langle\langle ij  \rangle\rangle,\sigma}( \hat{c}_{j \sigma}\d \hat{c}_{i \sigma}) \\
&+\Delta \sum_{i \in A}\hat{n}_i-\Delta \sum_{i\in B}\hat{n}_i -\mu \sum_{i, \sigma}\hat{c}_{i \sigma} \d \hat{c}_{i \sigma}+U\sum_{i}\hat{n}_{i \uparrow}\hat{n}_{i \downarrow}.  \label {model}
\end{split}
 \end{equation}
\noindent
Here $\hat{c}^\dagger_{i \sigma}$ creates electron with spin $\sigma$ on site $i$ of a 2-d square lattice which we regard as made of two square sublattices A and B whose sites are nearest neighbour of each other; $t$ is the nearest neighbour hopping, $\tp$ is the second neighbour hopping \cite{foot_note2}, $U$ denotes the on-site coulomb repulsion  and +$\Delta$(-$\Delta$) denotes the ``ionic'' potential for the A(B) sublattice.  The chemical potential ($\mu$) is chosen so that the average occupancy  $\left(\langle \hat{n}_A \rangle + \langle \hat{n}_B \rangle \right)/2  \equiv (n_A + n_B)/2$, is $1$  corresponding to the "half-filling" constraint. \\

%\section{HF Results}
\subsection{Simple limits of the IHM}
Some simple limits of the above model are relatively easy to understand. In the atomic limit ($t= \tp=0$), when $U/2 < \Delta$ the ground state of the system has two electrons on every B site and no electrons on the A sites, resulting in the maximal charge density wave order, and with a single particle excitation gap of $\Delta-U/2$. This is the atomic limit of the (correlated) BI phase. In the other case, when $U/2 > \Delta$, the ground state of the system is the (atomic limit of the ) Mott Insulator (MI) phase, with 1 electron at each site, and a  gap for single particle excitations equal to $U/2-\Delta$. Clearly at $U = 2\Delta$ the system has gapless single particle (electron or hole) excitations, and can of thought of as "metallic".\\
\\
For $U=0$ the model can be diagonalized exactly, in terms of two bands of electron creation and destruction operators with quasi-momentum or wave-vector labels $\bf{k} \equiv (k_x, k_y)$. The (spin independent) band dispersion relations in the full Brillouin Zone (BZ) are given by
\be
\begin{split}
\xi_{\bf{k}}^\pm& \equiv \epsilon^\pm_{\textbf{k}}-\mu \\
&= -4 \tp \cos{k_x}\cos{k_y} - \mu \pm \sqrt{ \Delta^2 + (2t(\cos{k_x} + \cos{k_y}))^2}.
\end{split}
\en
\noindent
When $\tp = 0$ the two bands are separated in energy by a band gap ($E_{gap}$) of   $2\Delta$ along the square contour  corresponding to $k_y = \pm \pi \pm k_x$. So at half filling the lower band is completely filled and the upper band is empty, resulting in a band insulator (BI). But when  $\tp \neq 0$, the eigenstates for $\textbf{K}=(0,\pm\pi),(\pm\pi,0)$ and $(\pm\pi/2, \pm\pi/2)$ are no longer degenerate. The bottom of the upper, conduction band  is $\epsilon^+_{\textbf{K}} =  \Delta$  at $\textbf{K}=(\pm\pi/2, \pm\pi/2)$ whereas the top of the lower, valence band is  $\epsilon^-_{\textbf{K}} = 4\tp - \Delta $ at  $\textbf{K}=(0,\pm\pi),(\pm\pi,0)$. Hence, for $\tp \neq 0$  and $\Delta > 2\tp$ the two bands have an indirect band gap $E_{gap}$ = $-4\tp + 2 \Delta$. As long as $E_{gap}>0$   the system (at half-filling) continues to be a Band Insulator (BI). As one increases $\tp$ from 0, the band gap monotonically decreases from $2\Delta$ and  become zero at $\tp=\Delta/2$, whence we get a BI to (band) metal transition.\\

\subsection{HF Mean Field Theory of the IHM}
\begin{figure}
 	\includegraphics[width=8cm]{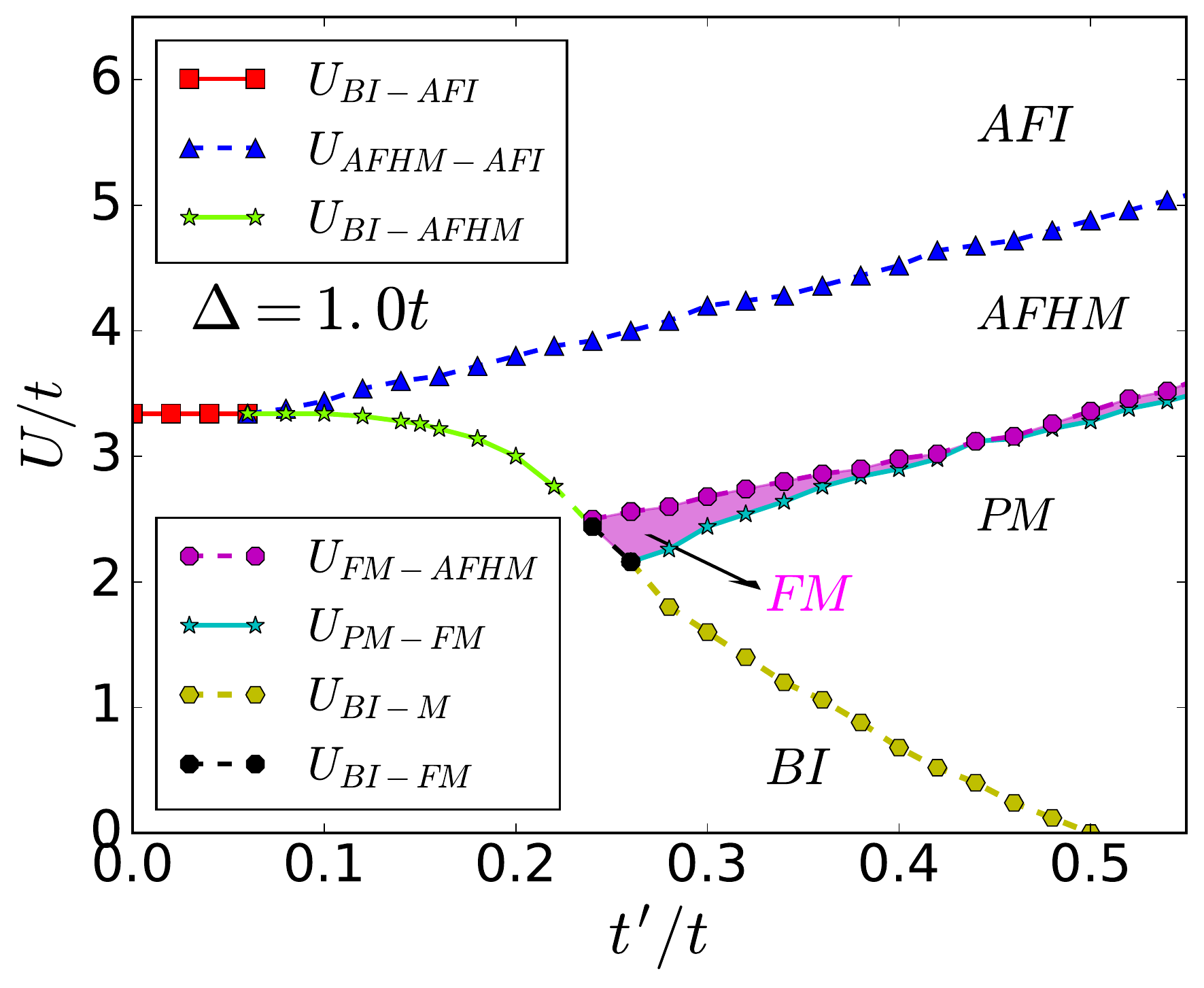}
 	\caption{Phase diagram of the $t-\tp$ IHM in the $\tp$-U plane at half filling using Hartree-Fock Mean Field Theory at zero temperature for $\Delta=1.0t$. %Different regions of the phase diagram are labelled by abbreviations characterizing the corresponding phases, such as BI $\equiv$ Band Insulator, PM $\equiv$ Parmamagnetic Metal, FM $\equiv$ Ferrimagnetic Metal, AFHM $\equiv$ Anti-Ferromagnetic Half Metal, AFI $\equiv$ Anti-Ferromagnetic (Mott) Insulator.   The lines in the phase diagram labelled with the format $U_{phase1-phase2}$ means that the system goes from "phase1" to "phase2" upon increasing $U$ across the line.  
For $\tp<0.06t$, the system undergoes a direct transition from BI to AFI phase with increase in $U$. For $0.08t <\tp < 0.24t$, the system goes from a BI phase to AFI phase via an intermediate AFHM phase upon increase of $U$, whereas for $0.24t<\tp<0.26t$, the system shows three  transitions, from  BI to FM, from FM to AFHM and eventually from AFHM to AFI as $U$ is increased. For $0.26t < \tp < 0.5t$, the PM phase intervenes between the BI and the other phases, whereas for $t^\prime>0.5 t$ there is no BI phase, and the transition is from the PM into other phases. Full lines in the phase diagram indicate first order transition lines, whereas dashed lines indicate continuous transitions. }
 	\label{HFphase}
\end{figure}

For  $U \ne 0$  the IHM is no longer exactly solvable. A simple approximate solution can be obtained using  the (unrestricted) Hartree-Fock mean field theory (HF-MFT), where, as described in detail in Appendix A, at the simplest level one approximates the Hamiltonian as an effective quadratic Hamiltonian allowing for mean field order parameters  corresponding {\em only} to the sublattice and spin resolved occupancies $\langle \hat{n}_{\alpha\sigma} \rangle  \equiv n_{\alpha\sigma}$, or, equivalently, to the staggered occupancy ($\delta n \equiv (n_B-n_A)/2$), the staggered magnetization ($m_s \equiv (m_A - m_B)/2$) and the uniform magnetization ($m_f \equiv (m_A + m_B)/2$) where $m_\alpha$ is the magnetization on any site belonging to sublattice $\alpha$ (see Appendix A for details). The resulting band dispersion relations, now spin dependent, are given by
\be
\begin{split}
&{\tilde{\xi}}^\pm_{\bf{k} \sigma} \equiv \tilde{\epsilon}^\pm_{\textbf{k} \sigma}- \mu = -4 \tp \cos{k_x} \cos{k_y}  +\frac{U}{2} -\frac{\sigma U m_f}{2}-\mu\\
& \pm \sqrt{[\Delta-U(\frac{\delta n + \sigma m_s}{2})]^2 + [2t(\cos{k_x} + \cos{k_y})]^2}
\end{split}
\en
\noindent
clearly one can interpret these results in terms of an effective spin dependent staggered potential $\tilde{\Delta}_{\sigma} = \Delta-U(\delta n + \sigma m_s)/2$ and an effective spin dependent uniform potential $ -\sigma U m_f / 2$. The effective band gap ($\tilde{E}_{gap,\sigma}$) for electrons with spin $\sigma$,  determined by the difference between the bottom of the conduction band ($\tilde{\epsilon}^+_{\textbf{k} \sigma}$) and the top of the valence band ($\tilde{\epsilon}^-_{\textbf{k} \sigma}$), similarly to the U=0 case, is $\tilde{E}_{gap,\sigma}$ = $-4\tp + 2 \tilde{\Delta}_{\sigma}$, which is interaction and spin dependent. The order parameters are then determined self consistently by populating these effective non-interacting bands as per Fermi-Dirac statistics. The details are provided in Appendix A. The phases are characterised both by the order parameters that are non-vanishing, and the nature of the band dispersions. Amazingly, the resulting phase diagram obtained within the HF approximation, depicted in Fig.~\ref{HFphase}, shows all the same phases as obtained using the much more sophisticated and accurate DMFT+CTQMC calculations, albeit for quantitatively different ranges of the parameter values. Since qualitative features of this phase diagram can be easily understood given the simplicity of the HF approximation, we discuss key aspects of these results next (while relegating the details to Appendix A), before presenting a discussion of the key aspects of our DMFT+CTQMC calculations.\\

\subsection{HF Phase Diagram of the IHM}
First consider the HF phase diagram (Fig.~\ref{HFphase}) in the paramagnetic regime. In this case the system does not have any magnetic order and hence both the staggered magnetisation, $m_s$, and the uniform magnetisation, $m_f$, are zero. This implies that $\tilde{\Delta}_{\uparrow} = \tilde{\Delta}_{\downarrow}$. This  spin symmetry is also reflected in the single particle excitation band gap $\tilde{E}_{gap,\sigma}$. The resulting phase is adiabatically connected to the noninteracting BI phase, hence we use the same label for it. As $U$ increases, the density difference between the two sublattices, $\delta n$, decreases because $U$ does not prefer holes or double occupancy. As an effect the gap in the single particle excitation spectrum, $\tilde{E}_{gap}$, decreases. One might expect that as one increases $U$ further,  $\tilde{E}_{gap}$ might go to zero at a certain $U$, whence one would get a band insulator to metal transition. On the contrary, when $\tp=0$, one finds  that the system {\em never} goes to a paramagnetic metallic state within the HF theory, i.e., $\tilde{E}_{gap}$ is nonzero for all $U$.   However, more accurate calculations using single site DMFT  do lead to a metallic state, as pointed out first by Garg et al \cite{ArtiPara}. Later work \cite{Byczuk,ArtiHM,soumen,2D4,2D3} showed that this metallic solution is not stable against AF ordering\cite{soumen}. As is clear from Fig.~\ref{HFphase}, this result continues to hold even when $\tp \ne 0$, as long as $\tp \stackrel{~}{<} 0.26t$ for $\Delta=1.0t$. {\em One of the main new results of the work presented in this article is that for sufficiently large $\tp$ (eg., $\tp \stackrel{~}{>} 0.26t$ for $\Delta =1.0t$ as in Fig.~\ref{HFphase}), $\tilde{E}_{gap}$ does go to zero within the paramagnetic sector, in both the HF and DMFT methods, leading to a robust paramagnetic BI to Paramagnetic metal (PM) transition, and a stable correlation-induced PM phase.} \\

\begin{figure}
\includegraphics[width=8cm]{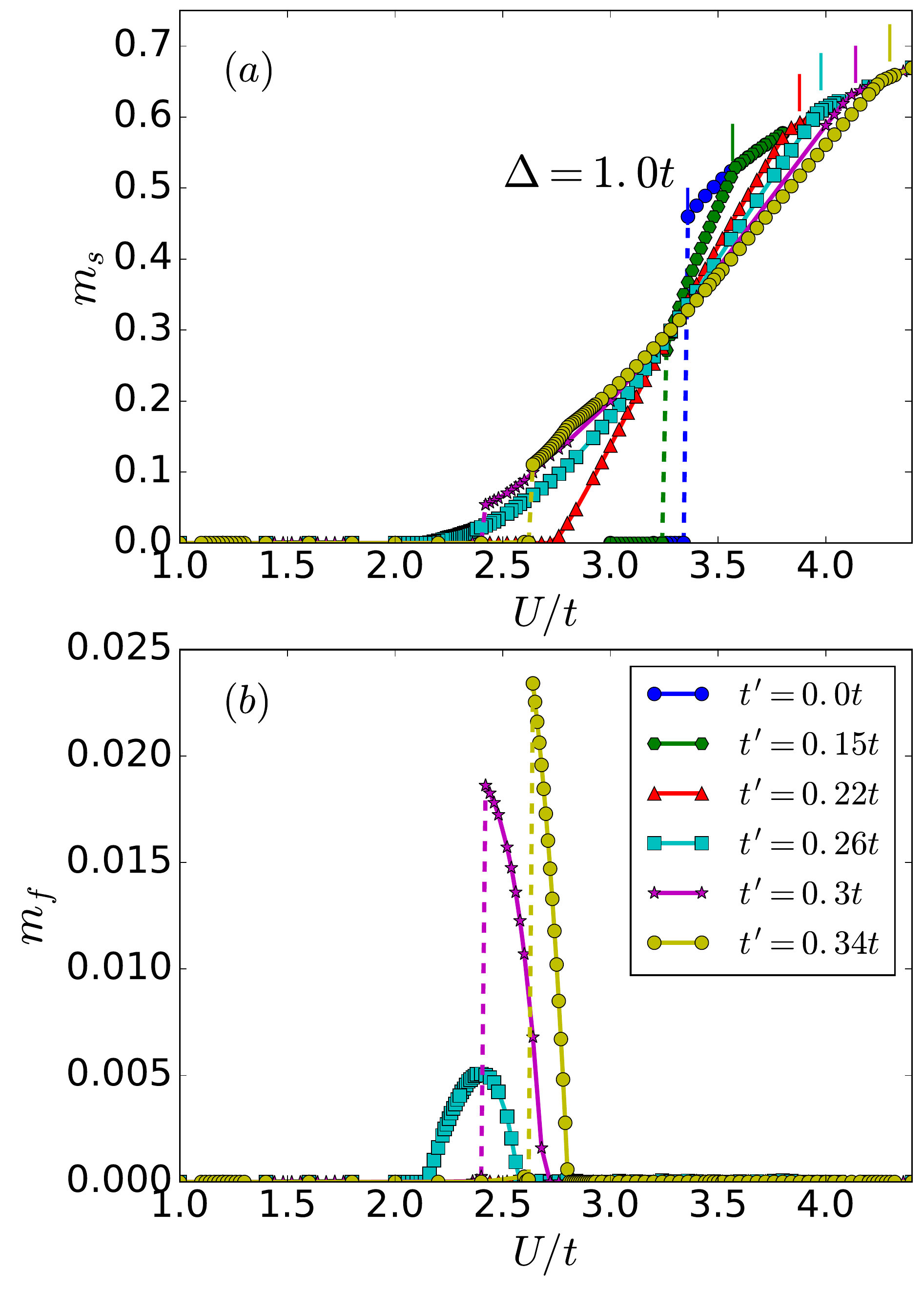}
 \caption{HF theory results for the staggered magnetization $m_s$  [panel(a)] and the uniform magnetisation $m_f$ [panel (b)] as  a function of the Hubbard interaction $U$ for different values of the second neighbour hopping ($\tp$) for $(n_A+n_B)$/2 = 1, $T=0$ and $\Delta=1.0t$. The magnetic transition is first order except for $\tp$ in the range from $0.215t$ to $0.26t$, where it is a continuous transition. Note that the values of $m_f$ are rather small, and exist only for relatively larger values of $\tp$ (for $\tp \ge 0.24t$), and only for limited ranges of $U$, corresponding to the magenta shaded region in Fig. \ref{HFphase}. Note that the FM-AFHM and the AFHM-AFI transitions are continuous, and show kinks (changes of slope) in the $m_s$ versus $U/t$ curves. }
\label{HFMS}
\end{figure}

Now we discuss results from the unrestricted HF theory where the spin symmetry breaking is allowed. Panels (a) and (b) of Fig.~\ref{HFMS} show the magnetic order parameters $m_s$ and $m_f$ at zero temperature within the HF approximation as functions of $U$ for different values of $\tp$. As $U$ increases, the staggered magnetisation $m_s$ turns on at a threshold value $U_c$, where $U_c$ is a function of $\tp$.
  %From results such as these we infer the threshold $U$, which we will refer to as  $U_c$ , above which magnetic order sets in. For the bulk of the region above $U_c$ , only $m_s$ is nonzero, except for the magenta shaded region in Fig.~\ref{HFphase}), where $m_f$ is also nonzero, although rather small, so the latter region is Ferrimagnetic.
At smaller $\tp$ values, the magnetic transition is a first order AF transition, with $m_f=0$ for all values of $U$. As $\tp$ increases, the discontinuity in $m_s$ across the transition decreases. We find that the transition is continuous for $0.215 t < \tp < 0.26 t$, and then again becomes first order as $\tp$ increases beyond $0.26t$, indicating the presence of multi-critical points at ($U=2.82t$, $\tp=0.215t$) and ($U=2.16t$, $\tp=0.26t$). The net magnetization $m_f$ seems to turn on for the first time for $\tp \sim 0.24t$ when the magnetic transition is continuous.  For larger values of $\tp$, $m_f$ also shows a first order jump indicating that the magnetic transition has become first order in nature.  The first order nature of the transition in $m_s$ as a function of  $U$ in Fig.~\ref{HFMS} is consistent with earlier results for $\tp =0$ on the  square lattice \cite{2D2,soumen}, and differ from the results in \cite{Byczuk}. Similar changes in the order of magnetic transition with increase in $\tp$ have been observed in the $t-\tp$ Hubbard Model  \cite{Vollhardt}. Note however, that the FM-AFHM and the AFHM-AFI transitions are continuous, and can be identified by kinks (changes of slope) in the $m_s$ versus $U/t$ curves. Clearly, these lead to other multi-critical points in the phase diagram.

We note that $U_c$ has an interesting  non-monotonic dependence on $\tp$. $U_c$ initially decreases with increasing $\tp$, reaching a minimum around $\tp=0.26t$, and then increases again. As we know from earlier studies on the IHM with $\tp = 0$ and bilayer systems, the value of $U_c$, at which the BI-AFI transition occurs, decreases with decreasing $\Delta$, the non interacting band gap\cite{soumen,Scalettar3}. Since the effective band gap, $\tilde{E}_{gap}$, decreases as $\tp$ increases, it is natural to expect that $U_c$ should decrease with increasing $\tp$. Another aspect of the next neighbour hopping is that $\tp$ introduces frustration against AFM ordering\cite{Vollhardt,duffy}, which effectively should increase the $U_c$ required for ordering. We believe that these two competing effects of $\tp$ are responsible for the observed  non-monotonic trend of $U_c$ with increasing $\tp$. This physics is seen in the phase diagrams obtained both within the HF and the DMFT methods.
\begin{figure}
	\includegraphics[width=8cm]{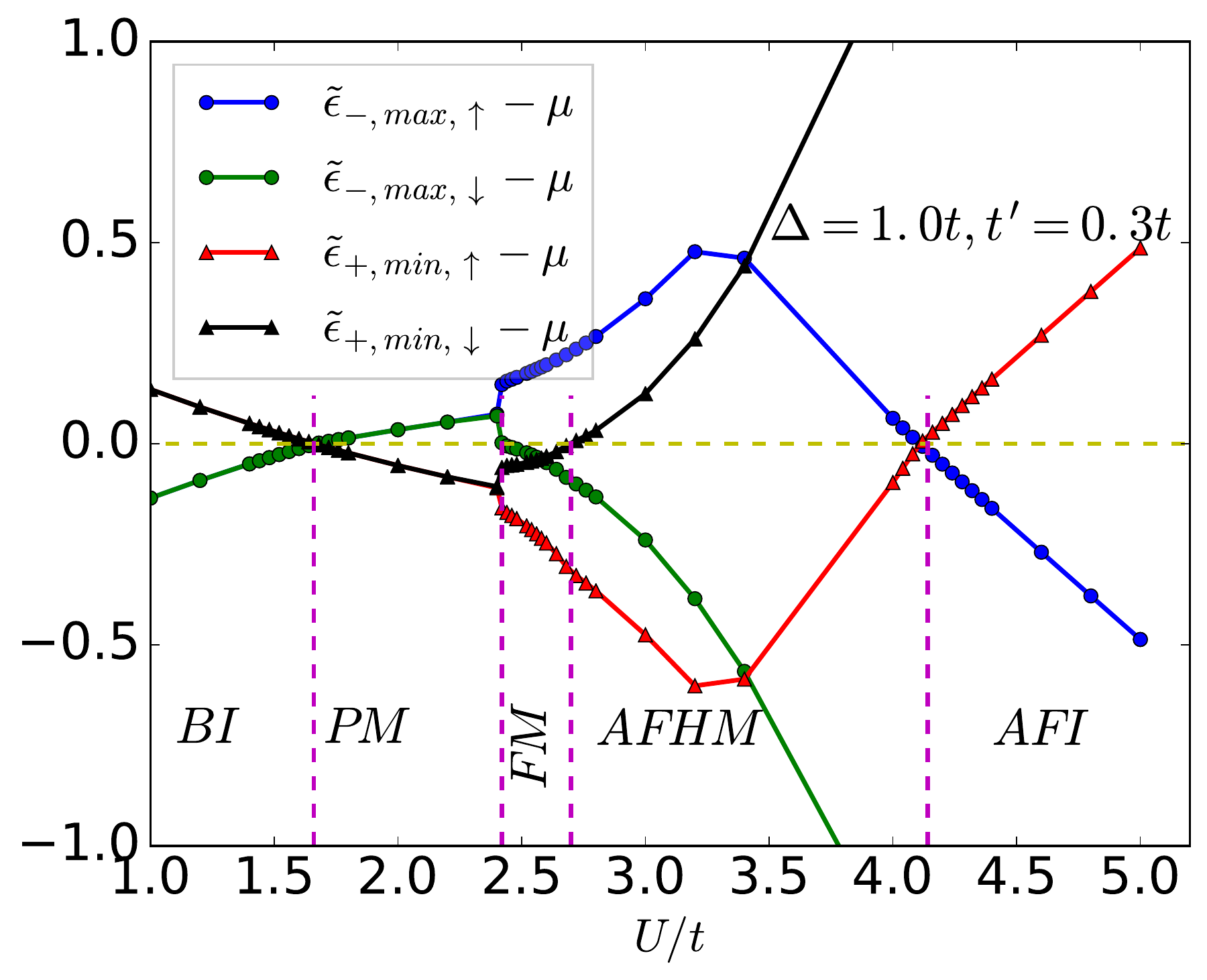}% in two colmn suze 6cm
	\caption{Maximum of the 'lower'(-) bands and the minimum of the 'upper'(+) band with respect to chemical potential ($\mu$) for both the spins plotted as functions of $U/t$ for $\tp = 0.3t$ }
	\label{band_min_max}
\end{figure}
Additional and interesting aspects of the magnetically ordered phases emerge from an examination of the self consistent HF band dispersions, and their gaps and overlaps. Fig.~\ref{band_min_max} shows the minimum of the upper (conduction) band, $(\tilde{\epsilon}^+_{min,\sigma})$, and the maximum of the lower (valence) band, $(\tilde{\epsilon}^-_{max,\sigma})$, measured relative to the chemical potential $\mu$ and for each spin channel, as a function of $U$ for $\tp=0.3t$. When the AF order parameter $m_s$ is nonzero, the effective staggered potential is spin dependent ($\tilde{\Delta}_{\uparrow} \neq \tilde{\Delta}_{\downarrow}$). In fact, for a positive $m_s$, corresponding to the A site being preferentially occupied by up spins,  the effective staggered potential for $\uparrow$ spins, ($\tilde{\Delta}_{\uparrow}$), is smaller than that for $\downarrow$ spins, ($\tilde{\Delta}_{\downarrow}$). So the effective band gap for $\uparrow$ spins, ($\tilde{E}_{gap,\uparrow}$) is less than that for $\downarrow$ spins, $\tilde{E}_{gap,\downarrow}$.
Because of this spin dependence of the effective band gap, if one dopes the system one gets a region of half metallic (HM) phase over a range of $U$ and doping\cite{ArtiHM}.

\begin{figure}
\includegraphics[width=9cm]{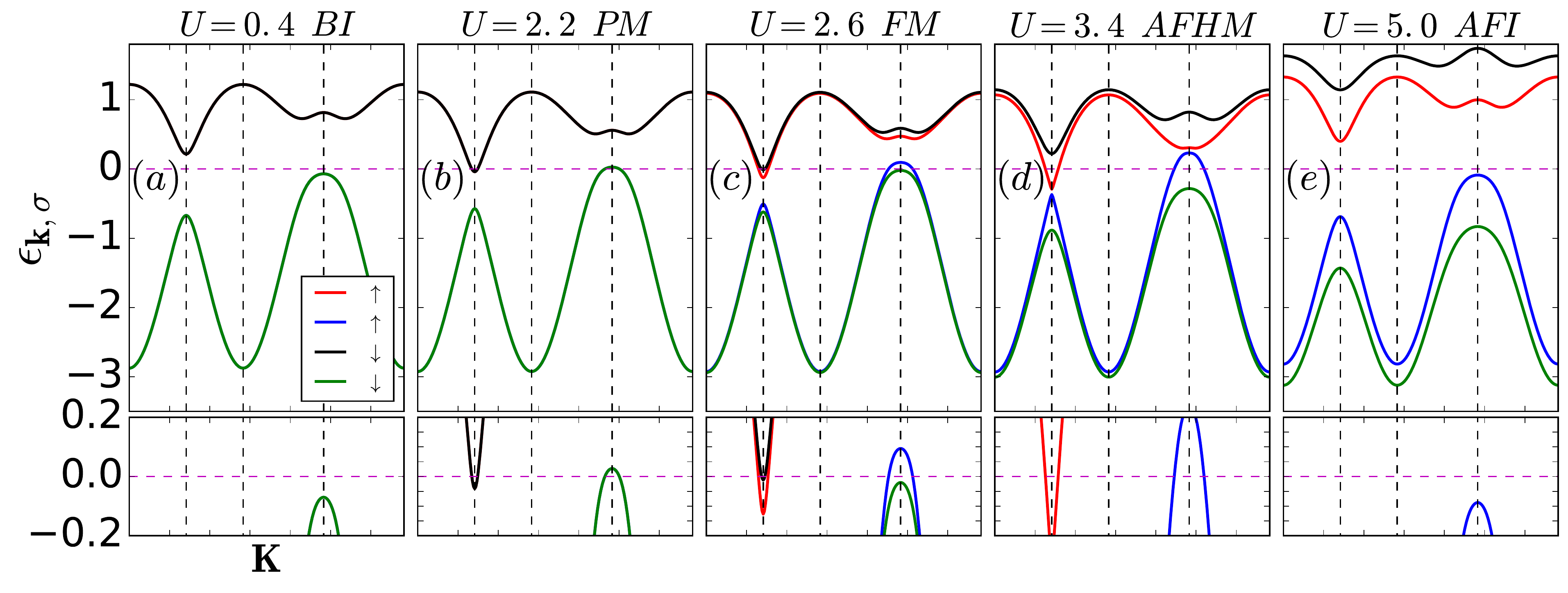}% in two colmn suze 6cm
\caption{Effective band dispersions, calculated using HF-MFT, plotted along the path in the Brillouin zone corresponding to $(0,0)\rightarrow(\pi/2,\pi/2)\rightarrow (\pi,\pi)\rightarrow (0,\pi)\rightarrow (0,0)$ (vertical doted lines indicate the break points) for different values of $U$ at $\Delta = 1.0t$ and $\tp=0.3t$. Zoomed in plots of the dispersions near zero frequency are presented in the lower panels. (a) For $U=0.4t$, the upper band has a minimum at $(\pi/2,\pi/2)$ and the lower band has a maximum at $(0,\pi)$. The chemical potential lies within the gap. Hence the system is in the (paramagnetic) BI phase. (b) For $U=2.2t$, both the upper and the lower bands cross the chemical potential. The upper band has an electron pocket at $(\pi/2,\pi/2)$ while the lower band has a hole pocket at $(0,\pi)$. This is the paramagnetic metallic phase. (c) For $U=2.6t$, the band structure is different for $\uparrow$ and $\downarrow$ spins. The upper bands  for both the spins cross the chemical potential, whereas only the lower $\downarrow$ spin band crosses the chemical potential. Hence we get a ferrimagnetic metallic state. (d) For $U=3.4t$, the $\uparrow$ spin bands cross the chemical potential whereas the $\downarrow$ bands are gapped out, corresponding to the AFHM state. (e) For $U=5.0t$, bands for both the spins are gapped out, and the system is in an AFI phase.}
\label{band}
\end{figure}

As illustrated in Fig.~\ref{band_min_max}, for sufficiently large values of ($\tp>0.26t$ for $\Delta=1.0t$), and for values of $U$ beyond a threshold, when the system is still spin symmetric, the upper and lower bands overlap for both the spin channels with the Fermi level being inside the bands, indicating a paramagnetic metal (PM) phase, with electron pockets near the conduction band minimum, and hole pockets near the valence band maximum. As $U$ increases further, the spin symmetry is broken. There is a small range in $U$ past the magnetic transition for which the down spin hole pocket disappears as its valence band is below the Fermi level while the down spin electron pocket and the up spin electron and hole pockets are still present. This corresponds to the ferrimagnetic metalic phase where $m_f\ne0$. %In fact in the Ferrimagnetic phase, as $m_f\ne 0$, at half filling the bands for both the spin channels {\em have to} cross the Fermi level, hence the Ferrimagnetic phase is necessarily metallic.
  On further increasing $U$, the bottom of the down spin conduction band goes above the Fermi level, and there is a broad range of $U$ where the $\uparrow$ spin gap is zero while the $\downarrow$ spin gap is finite. This parameter regime corresponds to an AF ordered half-metal (AFHM) phase.  As one increases $U$, $\delta n$ decreases but $m_s$ increases and at a certain $U$, $\delta n = m_s$. If one increases $U$ further,  $\tilde{E}_{gap,\uparrow}$ starts to increase from its negative value, and eventually it again becomes positive, resulting in a transition from the AFHM phase to an anti-ferromagnetic insulator (AFI) even within HF theory. Fig.~\ref{band} shows the band dispersions along specific paths (mentioned in the caption) in the Brillouin Zone for five representative values of $U$ chosen to correspond to all the five phases that arise for $\tp=0.3t$ as $U$ is increased, which again clearly brings out the above mentioned features of these phases. \\

\subsection{Single particle DOS within HF theory}

\begin{figure}
	\includegraphics[width=7.8cm]{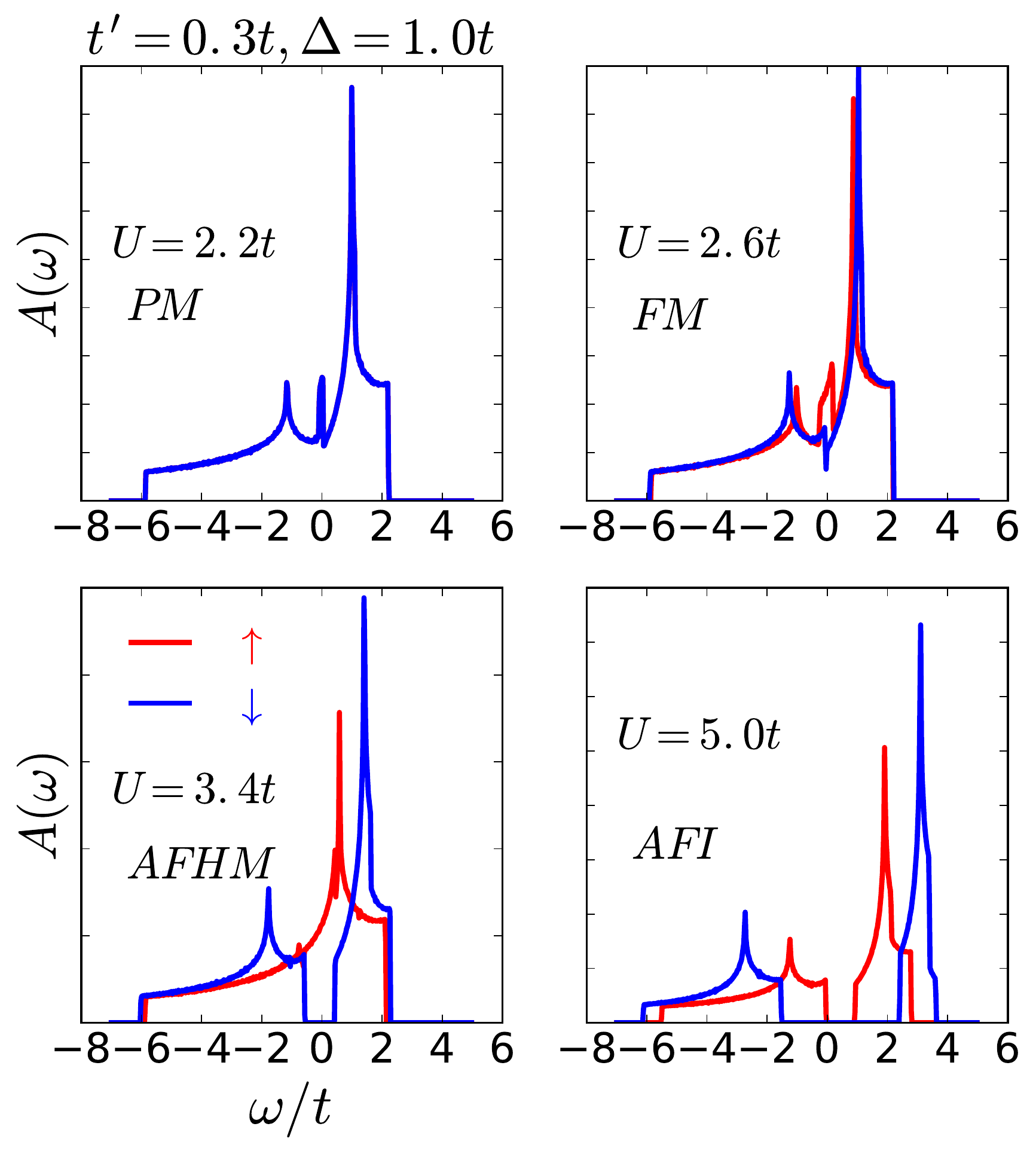}\\ 
	---------------------------------------------------------------------------------
	\includegraphics[width=7.8cm]{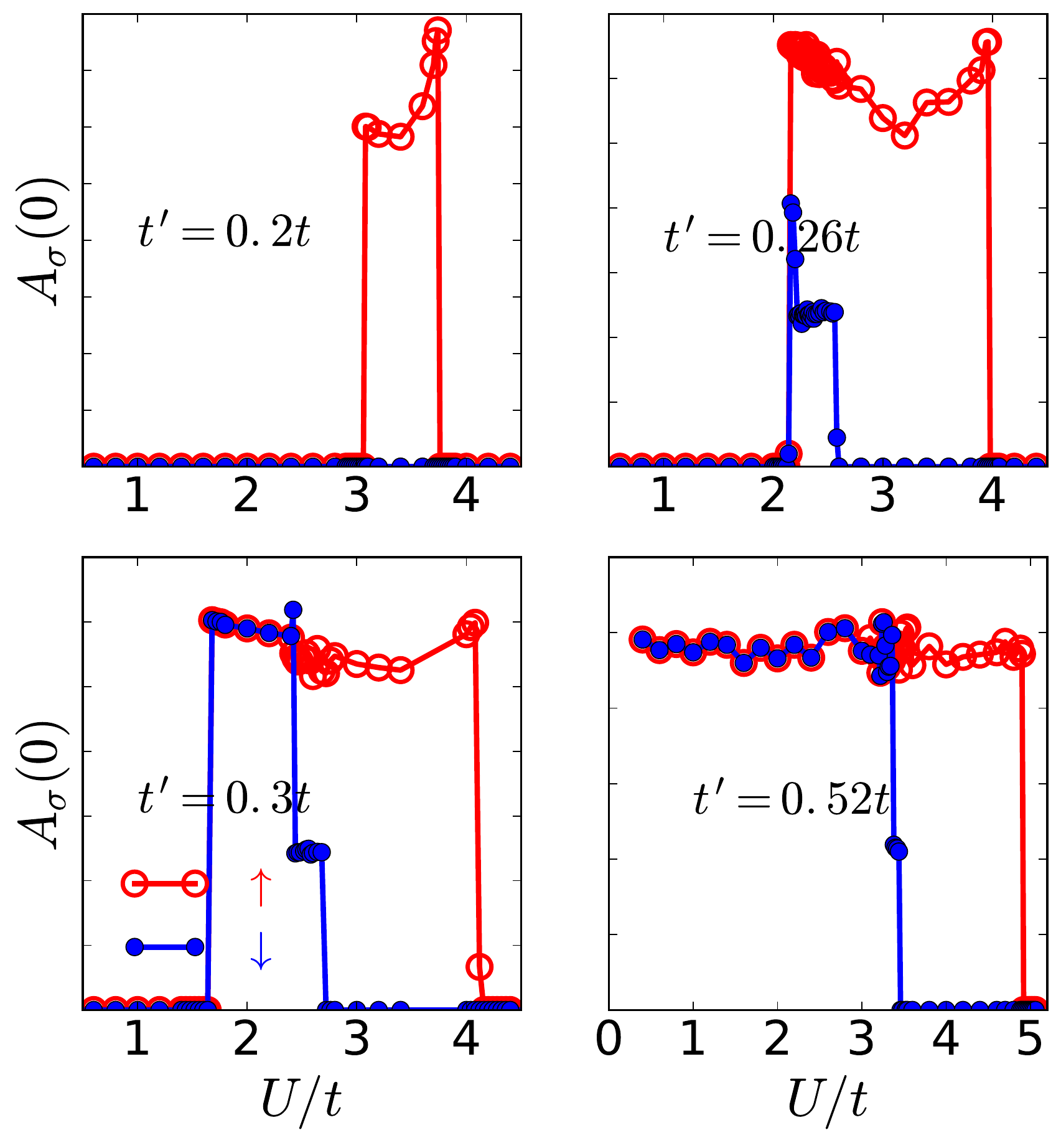}
	\caption{The top 4-panels show $A_\sigma(\omega)$ vs $\omega$ for $\tp=0.3t$ for a range of $U$ values. As $U$ increases one can see a transition from a PM BI to PM metal, which undergoes a transition into ferrimagnetic metallic and AF  half-metallic phase and eventually becomes an AFI. The bottom set of 4-panels present $A_\sigma(\omega=0)$ vs $U$ for a range of $\tp$ values.  }
	\label{HF_dos}
\end{figure}

All of the features of the phases and phase transitions discussed above can be seen more directly in the single particle density of states (DOS) for real frequencies, which can be calculated exactly within the HF approximation.
Specifically, Fig.~\ref{HF_dos} shows the results of our calculations for the spin dependent single particle DOS averaged over the two sublattices, $A_{\sigma}(\omega)= \frac{1}{2} \sum_\alpha A_{\alpha\sigma}(\omega)$, where
\be
A_{\alpha\sigma}(\omega) = -\frac{1}{\pi} Im~G_{\alpha\sigma}(\omega^+)
\label{spec-fn-def}
\en
is the single particle DOS for sublattice $\alpha$ and spin $\sigma$.
Here $G_{\alpha\sigma}(\omega)$ is the Green's function on sublattice $\alpha$, $\omega^+ = \omega+i\eta$ with $\eta \rightarrow 0$. The top panels in  Fig.~\ref{HF_dos} show the DOS for $\tp=0.3t$ and $\Delta=1.0t$ for the same values of $U$ as in Fig.~\ref{band}. For $U=2.2t$, in the PM phase, the system has spin symmetry (with $m_s=m_f=0$), the DOS for the two spin channels is the same and finite at the Fermi level (corresponding to $\omega=0$). For $U=2.6t$ in the FM phase, where both the magnetic order parameters are non-zero, the DOS is spin asymmetric, but finite at the Fermi level for both the spin channels. For  $U=3.4t$, $A_{\downarrow}(\omega)$ has a gap around the Fermi level, while $A_{\uparrow}(\omega=0)$ is finite. Because of the half-filling constraint, this necessarily implies that $m_f =0$; so this is the AFHM phase. For $U=5.0t$ gaps in the DOS are present in both the spin channels, corresponding to the AF (Mott) Insulating phase. The lower panels of Fig.~\ref{HF_dos} show the DOS at the Fermi level, $A_{\sigma}(\omega=0)$, as a function of $U$ for four illustrative values of $\tp$, bringing out the sequence of phases and phase transitions that occur as $U$ increases consistent with the complete phase diagram in Fig.~\ref{HFphase}. The existence of the BI (for smaller $U$ values) as well as AFI (for larger $U$ values) phases  where the DOS in both spin channels is zero, and the AF half-metal where $A_{\downarrow}(\omega=0) = 0$ but $A_{\uparrow}(\omega=0) \neq 0$, for different ranges of $U$ is clearly evident. For $\tp=0.3t$ the PM phase, where $A_{\uparrow}(\omega=0) = A_{\downarrow}(\omega=0) \neq 0$, and the FM phase, where $A_{\uparrow}(\omega=0) \neq A_{\downarrow}(\omega=0) \neq 0$, appear as well. For $\tp = 0.52 t$ the PM phase is present even at $U=0$, and there is no BI phase.

Although HF-MFT is very useful for the insights it provides into the nature of the various phases as discussed above, it is a mean field theory that neglects quantum fluctuations entirely. Hence one expects that it is likely to overestimate the regions of stability of the phases with broken symmetry. Needless to say, the inclusion of the quantum fluctuations, especially in the strong correlation regimes, pose considerable challenges, and in this paper we have restricted ourselves to doing this within the DMFT approximation, which correctly includes at least all the {\em local} quantum fluctuations, although the challenge of solving the resulting self-consistent impurity problem necessitates the further approximation of CTQMC. We discuss these DMFT calculations next.

%\section{Low Temperature Phase Diagram}
\section{DMFT Phase Diagram}
As mentioned earlier, the central result of our paper is the low temperature phase diagram of the half-filled IHM obtained within DMFT+CTQMC for the 2d square lattice {\em in the presence of a second neighbour hopping term}. Fig.~[\ref{phase_diag}] shows the phase diagram for $\Delta=1.0t$. Recalling the salient features of this phase diagram, we note that for $\tp<0.05t$, increasing $U$ leads to a first order transition from the paramagnetic BI to an AFI, similar to the $t'$ = 0 case.  However, as anticipated in our motivation for studying the $t-t'~ IHM$, for $t' > 0.05t$ the first transition  induced by increasing U  is a continuous transition from the BI phase into the correlation induced paramagnetic metal (PM) phase. As $U$ is increased further, two new phases not present in the $t^\prime = 0$ case (except at one value of U for each $\Delta$ or with doping\cite{soumen,ArtiHM}), namely the AFHM and the FM phases, intervene before the eventual transition into the AFI phase for large enough $U$. The AFHM phase is seen for a broad range of values of the second neighbour hopping, $0.05 t < t' < 0.46 t$, and abuts the AFI phase, whereas the FM phase is seen only for $0.12 t < t' < 0.36 t$ , and abuts the PM phase. A direct transition from the PM to the AFHM phase occurs for  $0.05 t < t' < 0.12 t$ and $0.36 t < t' < 0.46 t$, which is first order in nature. So is the PM to FM transition, except for the range $0.12 t < t' < 0.18 t$, where it is continuous. Both the FM-AFHM and the AFHM-AFI transitions are continuous in nature. For $0.46t< \tp < 0.5t$, the system undergoes only two transitions with increasing $U$, a continuous transition from the BI to the PM phase, followed by a first order transition into the AFI. For $\tp > 0.5t$ there is no BI phase. In this range of $\tp$, the system goes through  only one transition as $U$ increases, namely, a first order transition from the PM to the AFI phase. We note that indeed, as mentioned earlier, the values of various transition lines within the DMFT+CTQMC method are  very different from those obtained from HF theory. Due to quantum fluctuations captured within the DMFT+CTQMC method, the magnetic order sets in at larger values of $U$ as compared to the HF theory as expected. Furthermore, the width of the paramagnetic metal phase and the ferrimagnetic metal in $U-\tp$ plane is enhanced while the AF half-metal phase is diminished in width in the phase diagram from the DMFT+CTQMC method compared to that from the HF method.

In the rest of this section, we discuss the details of the DMFT results based on which this phase diagram has been constructed. While the identification of magnetic order is relatively easy in the DMFT method, it is much harder to distinguish insulating from metallic or half metallic behaviour, as the CTQMC produces imaginary time data, and precise spectral information in real frequency is therefore hard to generate. Although the frequency dependent spectral functions can not be obtained in a simple reliable manner from DMFT+CTQMC method, as we show later, one can look for features of metallic or insulating behaviour in momentum distribution function $n_{k\sigma}$, which is obtained by summing over Green's functions at imaginary frequencies. The insights gained from the HF method are also helpful in this regard. Furthermore, note that the use of the CTQMC methods constrains the calculations to be done at a non-zero temperature. The calculations presented in our paper are at $\beta t$=50.0, which we believe is at a low enough temperature that we expect our results to be close to the zero temperature results.
\begin{figure}
  \includegraphics[width=9cm]{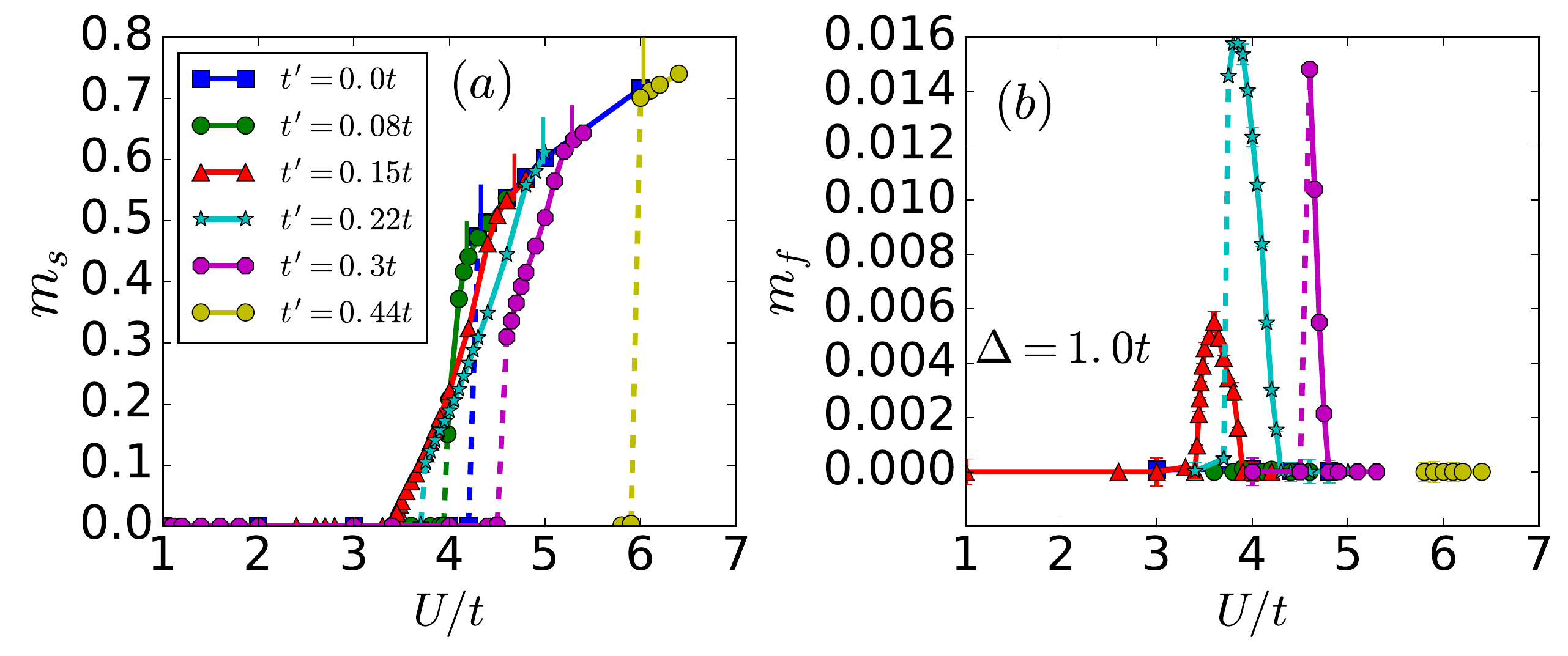} % two column size6
  \caption{The staggered magnetization $m_s$ and the uniform magnetization $m_f$ as a function of the Hubbard interaction $U$ for different values of the second neighbour hopping $\tp$ obtained from the DMFT calculations with $(n_A+n_B)/2 = 1$, $\beta t$=50.0 and $\Delta=1.0t$ held fixed. The discontinuity or jump in $m_s$ across the transition changes non-monotonically with $\tp$, similar to the HF results. For small values of $\tp$, the transition is strongly first order, but the jump decreases as one increases $\tp$ and vanishes around $\tp = 0.12t$. For $0.12t<\tp<0.18t$, the magnetization turns on with a continuous transition but on further increase in $\tp$, again $m_s$ and $m_f$ turn on with a jump at $U_c$. Thus the order of transition goes from first order at smaller $\tp$ to second order at intermediate $\tp$ to first order again at  lager $\tp$ indicating the presence of two multi-critical points in between along the $U_c$ versus $\tp$ line. The threshold value of interaction ($U_c$), where magnetic order sets in, initially decreases with increasing $\tp$, has its minimum value around $\tp = 0.18t \equiv t^{\prime\star}$, and again increases for larger $\tp$. %The vertical color coded indicator lines for each $\tp$ in the $m_s$ versus $U/t$) plot(panel a) represent the $U/t$ values for the final transition to the AFI phase. For intermediate $\tp$ (all cases in the figure except for $t' = 0. 0 t$  and $t^\prime = 0. 44 t$ ), where the transition to the AFI phase is from the AFHM phase, it shows up as a continuous transition, but with a slope discontinuity (within the limitations of our numerics).
 For $ 0.12 t< \tp< 0.36t$, there is finite regime  of $U/t$ where both $m_f$ and $m_s$ are nonzero, corresponding to the FM phase. (In making the figure we have set values of $m_f$ less than 0.001 to zero, because the error in $n$ is of the order of 0.001).  }
  \label{m}
\end{figure}

\subsection{Staggered density and Magnetization}
 The staggered magnetization $m_s$ and the uniform magnetization  $m_f$ calculated within the DMFT+CTQMC are shown as functions of $U/t$ in Fig.~\ref{m} for $\Delta=1.0t$ and inverse temperature $\beta~t=50.0$ for various values of $\tp$. For $\tp=0$ the AF order turns on with a strong first order transition at $U_c\sim 4.2t$. When $\tp$ is turned on, first the threshold value $U_c$ decreases, reaches a minimum around $\tp=0.18t\equiv t^{\prime\star}$, and then increases again with further increase in $\tp$. Furthermore, the discontinuity in $m_s$ at the first order transition decreases with increasing $\tp$, and the system undergoes a continuous transition for the range $0.12t<\tp<0.18t$, where both $m_s$ and $m_f$ turn on smoothly at $U_c$. For $\tp \ge 0.2t$, again the staggered magnetization and the uniform magnetization turn on with a jump.  For $\tp=0$, at half-filling the system has particle-hole symmetry leading to $m_A = -m_B$, whence $m_f = (m_A+m_B)/2 = 0$, i.e., the magnetic order is purely AF. When $\tp$ is turned on $m_f$ becomes non-zero due to the absence of particle-hole symmetry in the system, although it remains rather small for a range of intermediate values of $\tp$, as shown in panel (b) of Fig.~\ref{m} for $\Delta=1.0t$. For $\tp<0.12t$, $m_f$ is zero for all values of $U$. For $0.12t < \tp < 0.36t$, there is a small range $U_1 < U < U_2$ in which $m_f$ is non zero and the system has ferrimagnetic order, whereas for $U > U_2$ the system has pure AF order, and for $U < U_1$ the system is paramagnetic. For $\tp > 0.36t$, the uniform magnetization is again vanishingly small for all values of $U$. In this regime the system undergoes a transition from a para-magnetic phase to the AF phase and there is no ferrimagnetic phase.
 % Note that the value of $\tp$ below which $m_f$ is zero matches with $t^{\prime\star}$, where the transition point $U_c$ as a function of $\tp$ has its minimum.
\begin{figure}
	\begin{center}
  \includegraphics[width=8cm]{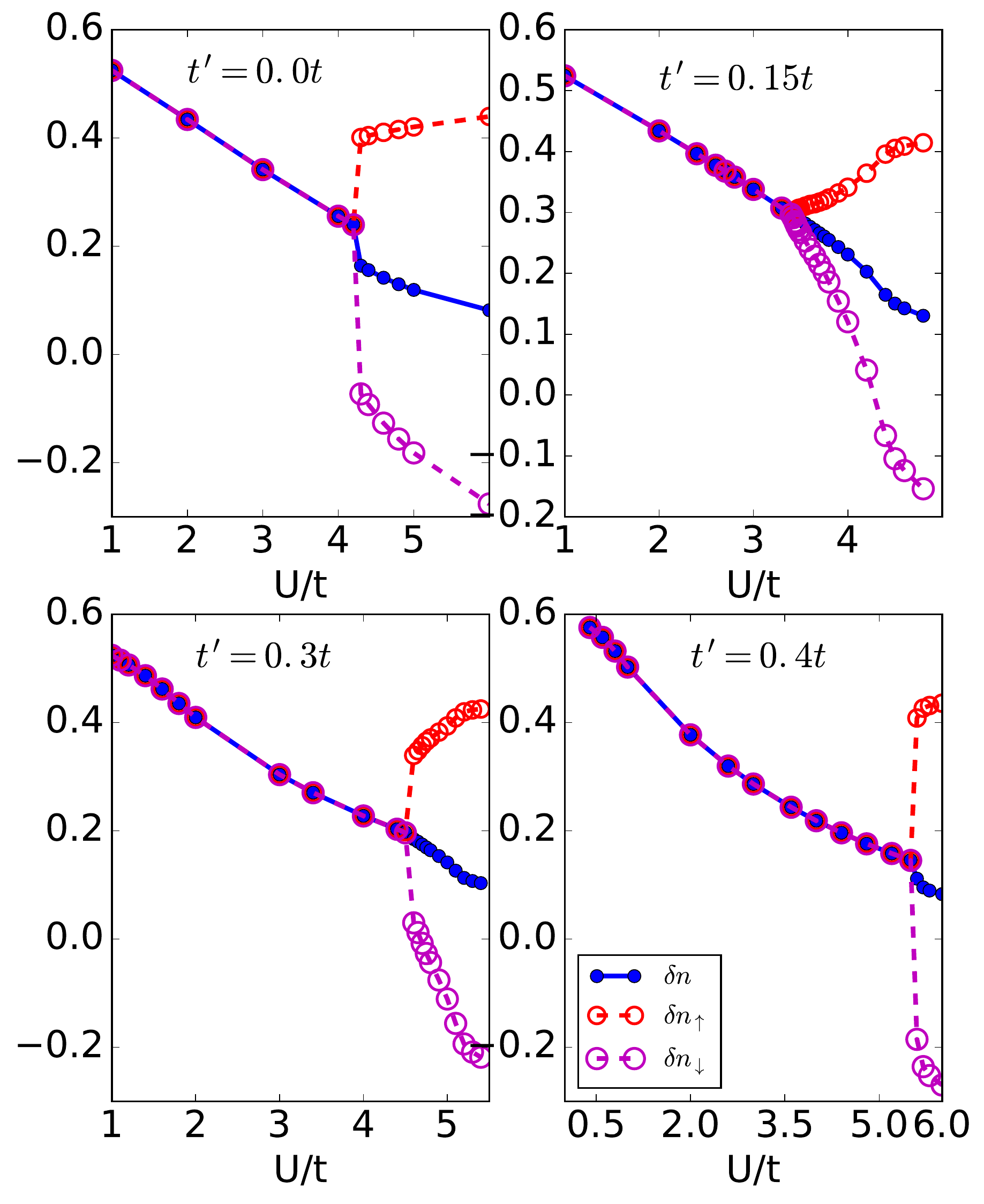}
  \caption{Spin resolved stagged occupancy, $\delta n_{\sigma}=(n_{B,\sigma}-n_{A,\sigma})/2$, and the mean staggered occupancy, $\delta n = 1/2\sum_{\sigma} \delta n_{\sigma}$, as a function of $U$ for different values of the second neighbour hopping ($\tp$) obtained using  DMFT+CTQMC. These results are for $(n_A+n_B)$/2 = 1 and $\beta t$=50.0, $\Delta=1.0t$. At $\tp=0.0t$ and for larger values of $\tp$, $\delta n_{\sigma}$ and $\delta n$  have first order jumps across the magnetic transition. However, at intermediate $\tp$ values $\delta n$ seems to change smoothly across the magnetic transition, though it exhibits slope changes at the continuous AFHM-AFI transition. }
  \label{dn_sigma}
\end{center}
\end{figure}
Fig.~\ref{dn_sigma} shows the spin resolved as well as the mean staggered densities given by ${\delta n}_\sigma = (n_{B\sigma}-n_{A\sigma})/2$ and $\delta n = ({\delta n}_\uparrow + {\delta n}_\downarrow)/2$ respectively, for inverse temperature $\beta t=50$ and $\Delta = 1.0t$. $\delta n$ is a decreasing function of $U$ for any value of $\tp$. An abrupt, first order drop in $\delta n_\sigma$ is seen across those values of $U_c$ where the magnetic transition is strongly first order, i.e., for small and large values of $\tp$. For intermediate values of $\tp$  where the magnetic transition (PM to FM transition) is continuous or weakly first order, $\delta n$ seems to change smoothly across the transition, but shows a slope change across the continuous AFHM to AFI transition.

\begin{figure}
    \includegraphics[width=9cm]{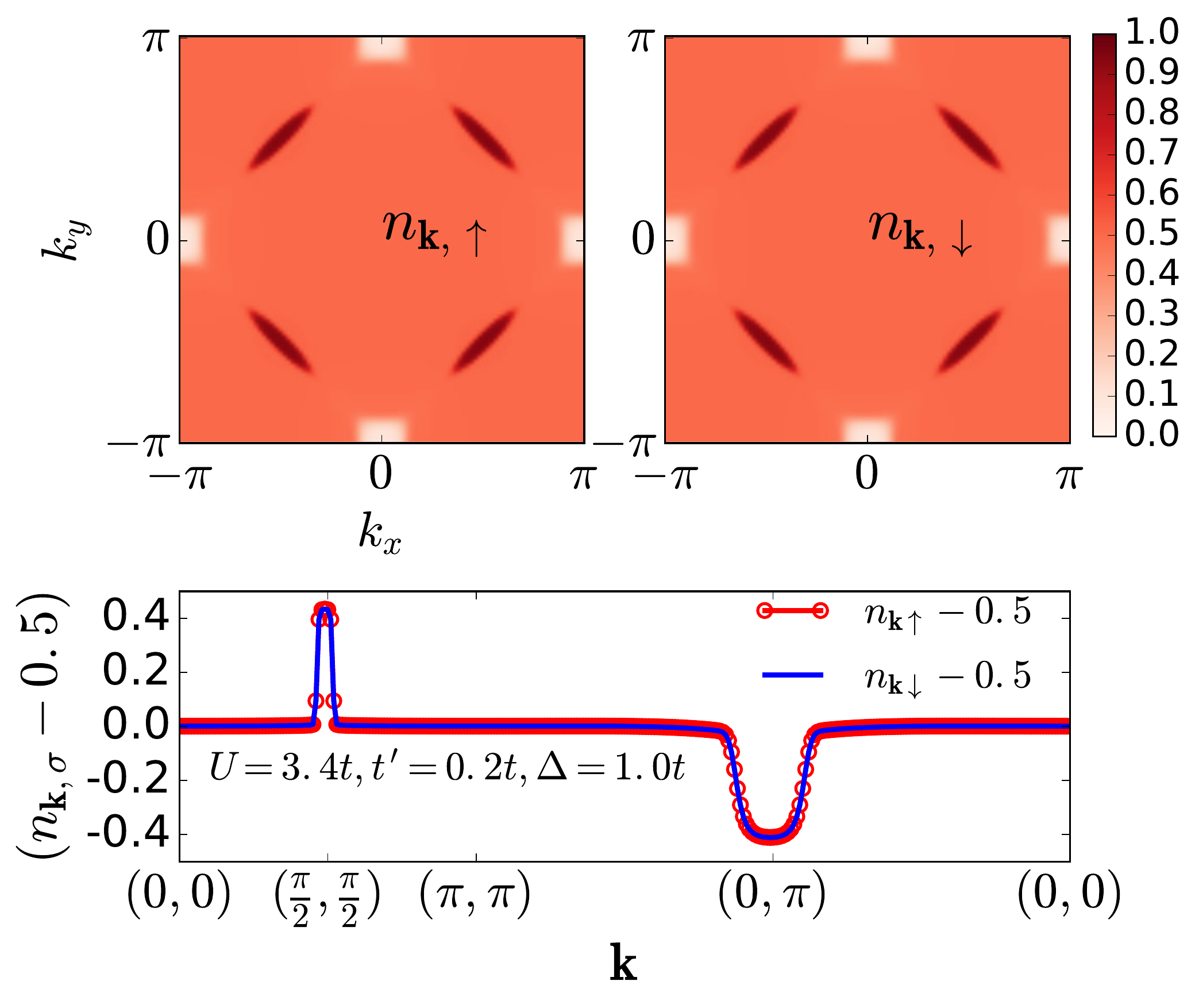}
 \caption{(a) False color plot of $n_{\textbf{k},\sigma}$ over the full Brillouin Zone (BZ) for $U=3.4t$ at half filling obtained using DMFT+CTQMC for $\beta t$=50.0, $\Delta=1.0t$. Around the symmetry points $(\pm \pi/2, \pm \pi/2)$ $n_{\textbf{k},\sigma}$ is close to the maximum value one corresponding to electron pocket where both the bands are occupied, and around the symmetry points $(\pm \pi,0)$ and $(0, \pm \pi)$, $n_{\textbf{k},\sigma}$ is close to zero corresponding to hole pockets, where both the bands are empty. (b) Numerical value of $(n_{\textbf{k},\sigma}-0.5)$ on a path along high symmetry directions in the BZ for $U=3.4t$ and $t^\prime=0.2t.$}
\label{nk_PM}
\end{figure}
\subsection{Momentum Distribution function}
 At zero temperature, whether a system is metallic or insulating is ideally characterized by the single particle density of states (DOS) or spectral function, calculated from the imaginary part of the single particle real frequency Greens function  (see equation \ref{spec-fn-def} and the related discussion). For a metal, the DOS at the Fermi energy is non zero, while for an insulator the DOS at the Fermi energy is zero. But within the CTQMC method, the Green's function is calculated only at imaginary times and at a finite temperature, and hence its Fourier transform is available only for imaginary Matsubara frequencies, $i\omega_n \equiv i (2n+1)\pi k_B T$ where n is an integer. In order to obtain the real frequency DOS at low frequencies close to the Fermi energy ($\omega=0$), an analytic continuation of the imaginary frequency Green's function is required. Perhaps the best known way of doing the analytic continuation is to use the maximum entropy method (MEM) but it has an in-built bias, involving an initial guess for the DOS, which makes it hard to get reliable real (low) frequency data to sharply distinguish between a metal and an insulator. As an alternative procedure that avoids analytic continuation altogether, in this paper we use the momentum distribution functions $n_{\alpha,\textbf{k},\sigma} = \int _{-\infty}^{0} d\omega \mathcal{A}_{\alpha\sigma}(\textbf{k},\omega)$ where $\mathcal {A}_{\alpha\sigma}(\textbf{k},\omega)$ is the {\em momentum resolved} spectral function for sublattice $\alpha$ and spin $\sigma$, to distinguish between the metallic and insulating nature of the system. Within the CTQMC $n_{\alpha,\textbf{k},\sigma}$ is calculated by doing a summation over the Green's function at Matsubara frequencies as
\be
n_{\alpha,\textbf{k},\sigma} = \frac{1}{\beta} \sum_{n} G_{\alpha,\sigma}(\textbf{k},i\omega_n) + \frac{1}{2}
\en
In a paramagnetic insulating phase at half filling, adiabatically connected with the noninteracting BI phase,  where the lower band is fully occupied while the upper band is vacant, with the chemical potential lying in the gapped region between the two bands, the spin-resolved momentum distribution function $n_{\textbf{k},\sigma} \equiv \frac{1}{2}\sum_\alpha n_{\alpha,\textbf{k},\sigma}$ will have a constant value of $1/2$ throughout the Brillouin zone (BZ). In a paramagnetic metallic state, adiabatically connected to the state in which the two (non-interacting) bands cross the Fermi level, we will have (interaction-renormalized) electron pockets in the conduction band leading to a larger (close to but less than 1.0) $n_{\textbf{k},\sigma}$ and hole pockets in the valence band, leading to a smaller (close to but greater than 0.0) $n_{\textbf{k},\sigma}$, in such a way as to keep the total particle number fixed at half filling ($(n_A+n_B)/2=1$). One can similarly use the spin dependence of the momentum distribution functions combined with the information on the magnetic order parameters to distinguish the FM and AFHM phases, as discussed below.
\begin{figure}
    \includegraphics[width=9cm]{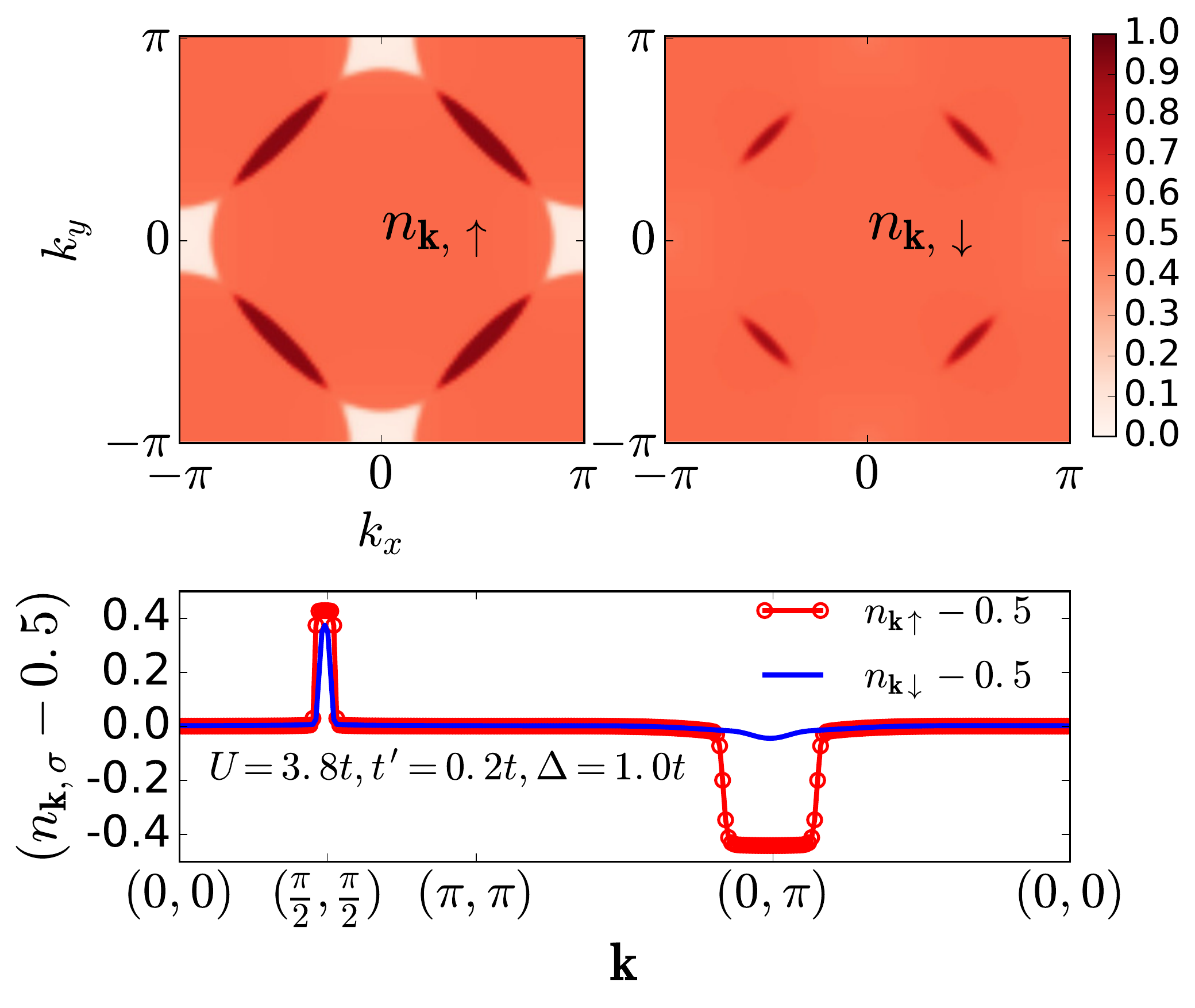}
 \caption{(a) False color plot of $n_{\textbf{k},\sigma}$ over the full BZ for $U=3.8t$ at half filling obtained using DMFT+CTQMC for $\beta t$=50.0, $\Delta=1.0t$.  For $\uparrow$ spin (top left panel) around $(\pm \pi/2,\pm \pi/2)$ we have electron pockets and around $(\pm \pi,0), (0, \pm \pi)$  we have hole pockets, and in the other $\textbf{k}$ regions the colors are uniform, corresponding to $n_{\textbf{k},\uparrow}$ of 0.5. In contrast, for $\downarrow$ spin(top right panel), there are much smaller electron pockets around $(\pm \pi/2,\pm \pi/2)$, and almost no hole pockets.  (b) Numerical values of $(n_{\textbf{k},\sigma}-0.5)$ on a path along high symmetry directions in the BZ for $U=3.8t$ with $t^\prime=0.2t.$}
\label{nk_FM}
\end{figure}

\begin{figure}
    \includegraphics[width=9cm]{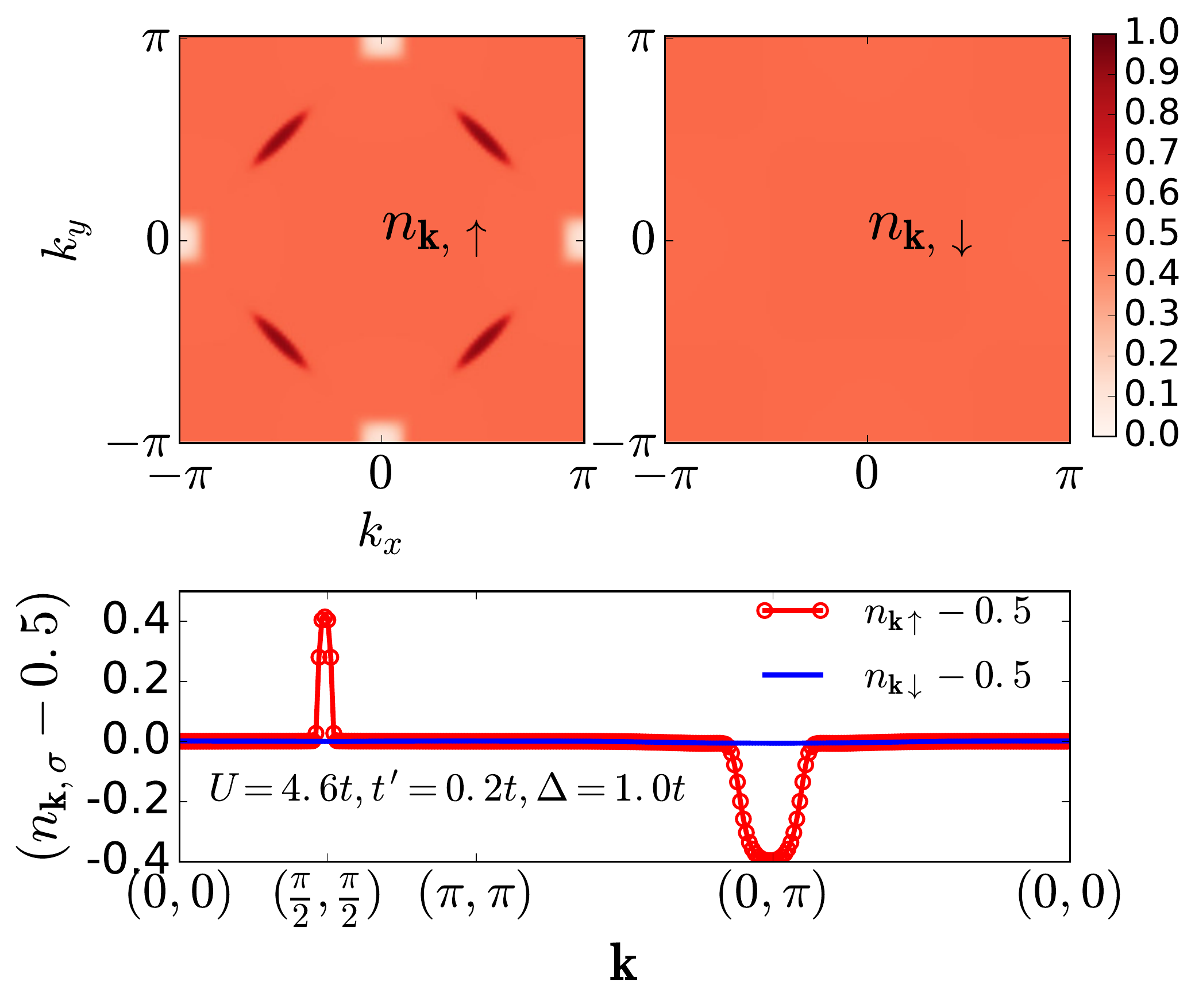}
\caption{(a) False color plot of $n_{\textbf{k},\sigma}$ over the full BZ for $U=4.6t$ at half filling obtained using DMFT+CTQMC for $\beta t$=50.0, $\Delta=1.0t$. For $\uparrow$ spin (top left panel) around $(\pm \pi/2,\pm \pi/2)$ we have electron pockets and around $(\pm \pi,0), (0, \pm \pi)$  we have hole pockets, and in the other $\textbf{k}$ regions the colors are uniform, corresponding to $n_{\textbf{k},\uparrow}$ of 0.5. In contrast, for $\downarrow$ spin(top right panel), there are no electron or hole pockets, and $n_{\textbf{k},\downarrow}$ is 0.5 over the entire BZ. (b) Numerical values of $(n_{\textbf{k},\sigma} - 0.5)$ on a path along high symmetry directions in the BZ for $U=4.6t$}  \label{nk_AFHM}
\end{figure}

\begin{figure}
\includegraphics[width=7.5cm]{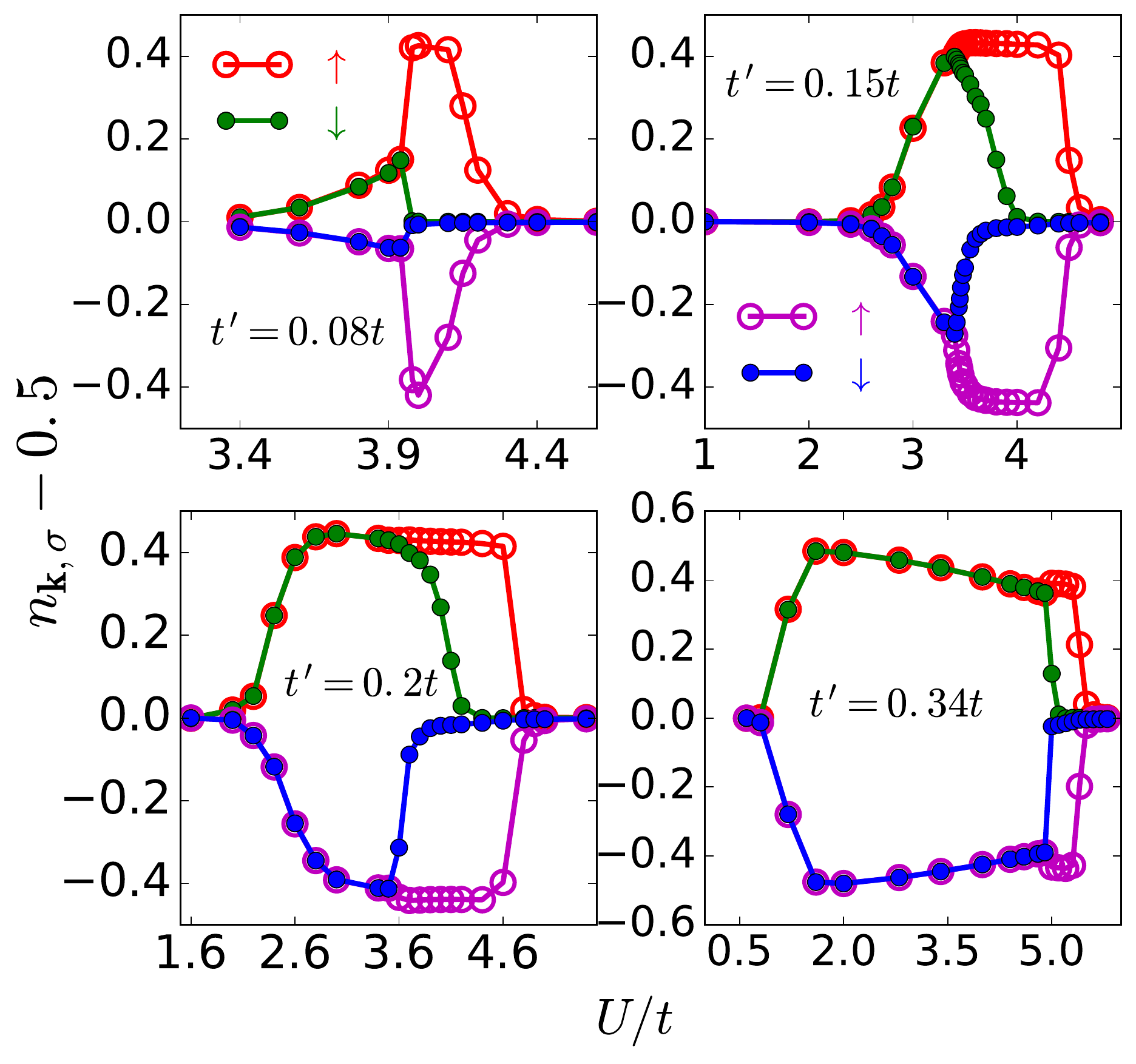}\\
------------------------------------------------------------------------------
\includegraphics[width=7.5cm]{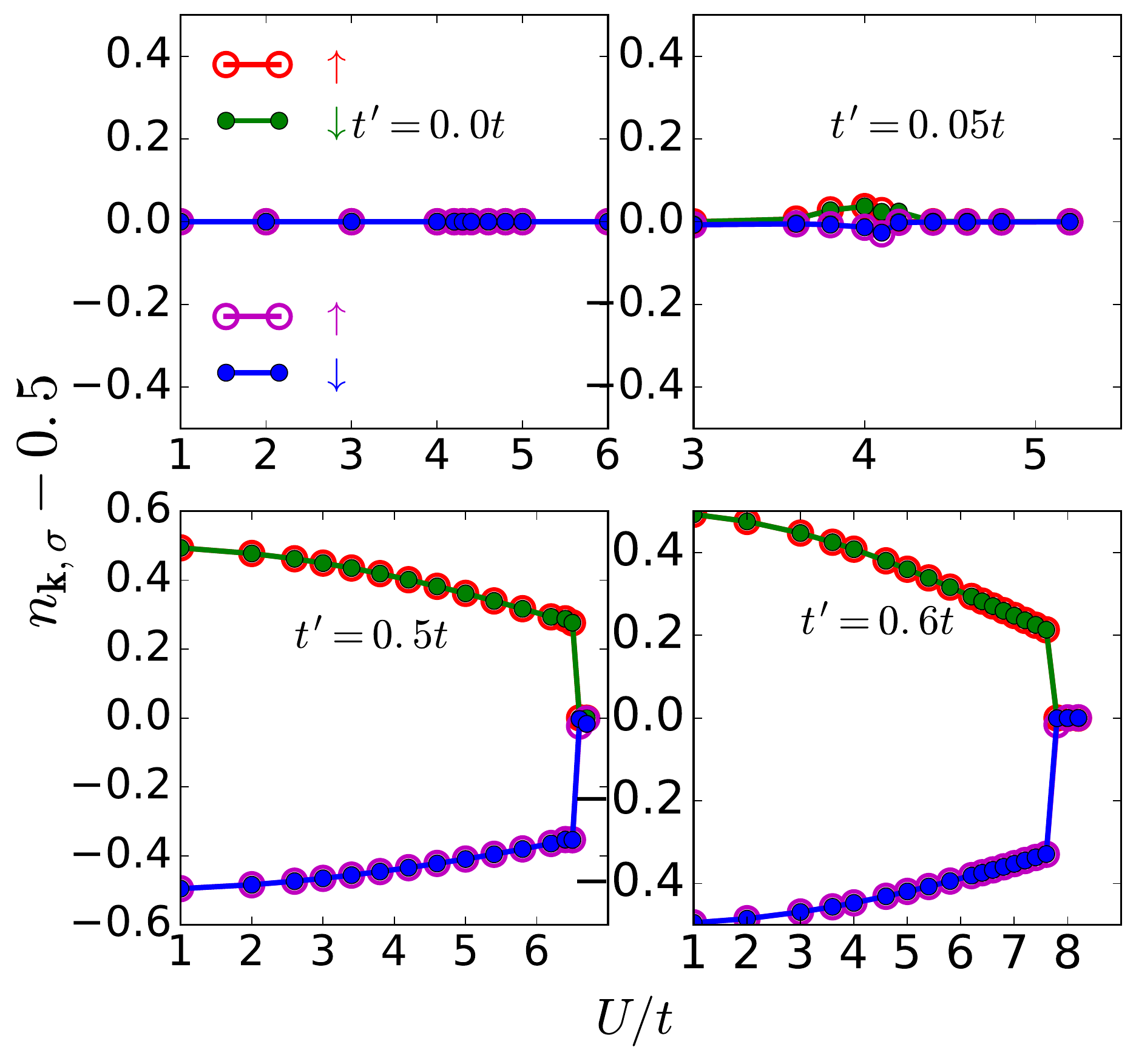}
\caption{($n_{\textbf{k},\sigma}-1/2$) at $\textbf{k} = (\pi/2,\pi/2)$ (red and green curves)  and $\textbf{k} = (0,\pi)$ (magenta and blue curves) vs $U$ for eight different values of $ \tp=0.08t, 0.1t, 0.2t, 0.34t$ (top set) and $\tp=0.0t,0.05t, 0.25t, 0.6t $ (bottom set).
%Where colored spin {$\color{red}\uparrow $}, {$\color{green}\downarrow $} for $n_{\textbf{k},\sigma}-1/2$ at $\vec{k}=(\pi/2,\pi/2)$ and {$\color{magenta}\uparrow $}, {$\color{blue}\downarrow $} at $\vec{k}=(0,\pi)$. %
In the metallic phases  ($n_{\textbf{k},\sigma}-1/2$) is  greater than zero at $\textbf{k}=(\pi/2,\pi/2)$, suggesting electron pockets, and is less than zero at $\textbf{k} = (0,\pi)$, suggesting hole pockets. For $\tp=0.2$  one  can clearly see the transition from the BI to the PM phase, followed by a transition into the FM phase, where $n_{\textbf{k},\sigma}$ becomes different for $\uparrow$ and $\downarrow$ spins. This is followed by a transition into an AF half-metal where  $(n_{\textbf{k},\sigma}-1/2)$ is different from zero for only one spin channel. Finally at large $U$, $n_{\textbf{k},\sigma} = 1/2$ for both the spin channels, corresponding to the AFI phase. Other panels similarly show the phases and transitions for other values of  $\tp$, and also how the width in $U$ of the different phases changes with $\tp$.
%Right side four panel: average of the values of ($n_{\textbf{k},\sigma}-1/2$) at $\vec{k}=(\pi/2,\pi/2)$   and $\vec{k}=(0,\pi)$ vs $U$ for four different $ \tp=0.08, 0.1t, 0.2t, 0.34t$.%
}
\label{nk_peak}
\end{figure}

 Fig.s~\ref{nk_PM}, \ref{nk_FM} and \ref{nk_AFHM} show false color plots of $n_{\textbf{k},\sigma}$ over the full BZ for $\tp$=0.2t for three values of $U$ at $\Delta=1.0t$. For $U=3.4t$ (Fig.~\ref{nk_PM}), the momentum distribution function is spin symmetric and has clearly visible renormalized electron and hole pockets as mentioned above, reflected respectively in the peaks at $\textbf{k} = (\pm \pi/2,\pm \pi/2)$  and the dips at $\textbf{k} = (0,\pm \pi), (\pm \pi, 0)$ in $n_{\textbf{k},\sigma}$, corresponding to the PM phase. Note also the particle-hole asymmetry because of the presence of $\tp$. As $U$ is increased and the magnetic order sets in, there develops an asymmetry in $(n_{\textbf{k},\sigma}-0.5)$ for the $\uparrow$ and $\downarrow$ spin channels, as shown in Fig.~\ref{nk_FM} for $U=3.8t$, with the $\uparrow$ spin channel having bigger electron and hole pockets, and also a larger net occupancy, compared to the $\downarrow$ spin channel. This is clearly a FM phase. On further increasing $U$, as shown  in Fig.~\ref{nk_AFHM} for $U=4.6t$, only $n_{\textbf{k},\uparrow}$ has Fermi pockets while $n_{\textbf{k},\downarrow}$ has the uniform value of $0.5$ everywhere in the BZ. Thus the down spin channel is insulating with a finite band gap while the up spin channel is a metal with both conduction and valence bands crossing the Fermi level, while the net occupancy in each channel is the same. This clearly corresponds to the AFHM phase of the system.
The complete picture about the multiple metal-insulator transitions and crossovers in the IHM, shown in the phase diagram of Fig.~\ref{phase_diag} requires the momentum distribution functions calculated within DMFT+CTQMC as illustrated above. A quick way of obtaining this picture is to study the variation in $n_{\textbf{k},\sigma}$ at just the centers of the electron and hole pockets as functions of the parameters of the IHM.  Fig.~\ref{nk_peak} shows  $(n_{\textbf{k}\sigma}-\frac{1}{2})$ at $\textbf{k} = (\pi/2,\pi/2)$ (red and green curves), and $(\pi,0)$ (magenta and blue curves) as a function of $U$ for several different representative values of  $\tp$, chosen to show the different sequences of phases and phase transitions that are possible. For example, consider the panel for $\tp=0.2 t$, for which  all five phases that we have discussed exist for different ranges of $U$, separated by four phase transitions (see Fig.~\ref{phase_diag}). In the insulating phases $(n_{\textbf{k},\sigma}-1/2)$ is zero and the onset of metallicity is indicated by $(n_{\textbf{k},\sigma}-1/2)$ deviating from zero. One can see that the first transition is continuous and from the BI to the PM phase, with spin symmetric occupancies, across the blue curve in the phase diagram of Fig.~\ref{phase_diag}. This is followed by the second transition, into the FM phase where {\em all} the $(n_{\textbf{k},\sigma}-1/2)$ are still nonzero but become different for $\uparrow$ and $\downarrow$ spins. The third transition from the FM to the AF half-metal, where $(n_{\textbf{k},\sigma}-1/2)$ is different from zero for only one spin channel, whereas the other spin channel is insulating. The final transition at large $U$, is a first order transition to the AFI phase, with $n_{\textbf{k},\sigma}=1/2$ again for both the spin channels. One can similarly see that the occupancy deviations from 0.5 shown in the other panels are consistent with Fig.~\ref{phase_diag}.
%The right side four panels in figure in Fig.~\ref{nk_peak} show the average $n_{\pi/2,\pi/2}-n_{\pi,0}$ vs $U$ for the same values of $\tp$. %
Note that a sharp distinction between the metal and the insulator can be made strictly only at $T=0$ based on whether the spectral functions have a gap around the Fermi level or not, which is reflected in the distribution function $n_{\textbf{k},\sigma}$ as well. However, CTQMC calculations are possible only at finite temperatures; therefore, even at the lowest possible temperatures accessible numerically, one can clearly distinguish between metallic and insulating phases only if the gap in the insulating phase is much larger than $T$. Close to the metal-insulator transition point, where the gap becomes very small, numerically it becomes very hard to locate the transition, as it is really a crossover. But within our numerical accuracy, we can still clearly see the presence of a correlation induced paramagnetic metallic phase, ferrimagnetic metallic phase and AF half-metal in the IHM, which is one of the main achievements of this work.

\section{Concluding Comments}
In conclusion, we have presented a detailed phase diagram of the half-filled IHM in the presence of a second neighbour hopping $\tp$ within DMFT+CTQMC. We have demonstrated that in this simple model of a band insulators (BI), due to the frustration of the anti-ferromagnetic (AF) order induced by $\tp$, one can stabilize a correlation induced paramegnetic metallic (PM) phase intervening between two insulating phases, namely the BI and the AF- Mott Insulator (AFI) for relatively large values of $\tp/t$, whereas the PM phase occurs as a precursor to an AF half-metallic phase, or a ferimagnetic metallic phase for smaller values of $\tp/t$. We believe that it is interesting how e-e interactions can dynamically close the gap in the band insulator and result in the formation of stable metallic phases which can be paramagnetic or ferrimagnetic, and even more interestingly, anti-ferromagnetic {\em half-metallic}.

The question of a whether there can be such a stable correlation induced  metallic phase intervening between the BI and the AF Mott Insulator phases in the half-filled IHM has been floating around in the community for almost a decade and had not been answered to the full extent earlier. In this work, we have used the momentum distribution function calculated within the DMFT+CTQMC for the analysis of metal-insulator transition, which, we think, is more reliable then using analytically continued single particle DOS. Our work presents a clean and complete answer to the above question by analysing the IHM at half-filling in the presence of a second neighbour hopping, and furthermore, has uncovered  a rich set of additional phases and phase transitions not known hitherto. We showed that, while for $\tp$ below a threshold $t_1$ an increase in $U$ induces a BI to AFI transition as in the $\tp=0$ case,  there is a range $t_1 < \tp  < t_2$ where the correlation induced PM phase and then a new AFHM phase intervene between the BI and the AFI phases as U increases.
In the next range of values of $\tp/t$, $t_2 < \tp < t_3$, the system undergoes four transitions involving five phases as the strength of $U$ is increased. First the paramagnetic BI undergoes a continuous transition into a PM, which at $U_c$ becomes a ferrimagnetic metal (FM) with the onset of uniform as well as anti-ferromagnetic order. As $U$ is increased further, a gap opens up in the single particle excitation spectrum of one of the spin channels while the other spin channel remains gapless, resulting in the formation of a half-metal which has pure AF order (the AFHM phase). On further increasing $U$, both the spin channels attain gaps in their single particle excitation spectra and the system becomes an AFI. For larger values of $\tp$, $t_3 < \tp < t_4$, the FM phase does not appear, and the transition from the BI to the paramagnetic metal is followed by a transition into the AFHM phase which eventually transforms to an AFI as $U$ increases. For even larger values of $\tp$, $t_4 < \tp < t_5$, the system undergoes only two transitions as $U$ increases with a broad PM phase intervening between the BI and the AFI phases. For $\tp > t_5$ there is no BI phase, and there is only one transition, between the PM and the AFI phases. The sizes and the end points of the different ranges of $\tp$ mentioned above vary with $\Delta$, and we hope to study their systematics in future work.

Furthermore, in our work we also showed that a similar phase diagram can be obtained in a simple unrestricted HF theory,  although within the simple HF theory the region of stability of the PM phase is suppressed compared to the DMFT results due to two reasons, both arising presumably from the missing quantum fluctuation in the HF calculations. First, the transition values $U_c$ for the onset of magnetic order are lower in the HF theory in comparison to DMFT; second, the AFM half-metal phase is too broad in the HF theory. Within the HF theory, unlike in the DMFT results, once $\tp$ is large enough to stabilize the correlation induced PM phase, the AFHM phase seems to always intervene between the PM and AFI phases, and a direct transition from the PM to the AFI is never seen.

It will definitely be very interesting to investigate the existence of these correlation driven metallic and half-metallic phases as well as the nature of various phase transitions involved, experimentally in real materials as well as in cold atom experiments where IHM has already been simulated on a 2d honeycomb lattice~\cite{UCA1}, and theoretically, using the more sophisticated and computationally demanding cluster DMFT and other methods\cite{final_note}. We hope very much that our work presented here stimulates further research in these directions.

\appendix
%\small{\input{intro/appendix}}
\section{Hartee-Fock theory}
In the HF approximation we break up the interaction term within a mean-field approximation as
\be
U \hat{n}_{i \uparrow} \hat{n}_{i \downarrow} \approxeq U[\langle \hat{n}_{i \uparrow} \rangle \hat{n}_{i \downarrow}+ \hat{n}_{i \uparrow} \langle \hat{n}_{i \downarrow} \rangle - \langle \hat{n}_{i \uparrow} \rangle \langle \hat{n}_{i \downarrow} \rangle],
\en
Furthermore, in this work we make the simplifying ansatz that $\langle \hat{n}_{i \sigma} \rangle$ is the same, $n_{A\sigma}$ or $n_{B\sigma}$, for all A or B sites respectively. Transforming to  creation (and destruction) operators labelled with wave vectors $\textbf{k}$ according to
\be
\hat{c}_{\alpha \textbf{k} \sigma}\d \equiv \sum_{j \in \alpha} e^{i \textbf{k} \cdot {\textbf{r}}_{j \alpha}} {\hat{c}}_{j \alpha  \sigma}\d
\en
where  $\alpha = A, B$ is the sub-lattice index, we can write the (operator part of the) HF hamiltonian as
\be
\begin{split}
	\hat{H}_{IHM; HF}& =  \sum_{\textbf{k},\sigma} \{\epsilon_{\textbf{k}}( \hat{c}_{A \textbf{k}  \sigma} \d \hat{c}_{B \textbf{k} \sigma}+ H.c.) \\
	&+  A_{\textbf{k}  \sigma} \hat{c}_{A \textbf{k} \sigma} \d \hat{c}_{A \textbf{k} \sigma} +  B_{\textbf{k}\sigma} \hat{c}_{B\textbf{k} \sigma} \d \hat{c}_{B\textbf{k} \sigma} \}
\end{split}
\en
where $A_{\textbf{k}  \sigma} \equiv \Delta -\mu + \epsilon^\prime_{\textbf{k}}  + U n_{A\bar{\sigma}} $ and $B_{\textbf{k}\sigma} \equiv -\Delta -\mu + \epsilon^\prime_{\textbf{k}} + U n_{B\bar{\sigma}} $, with  $ \epsilon^\prime_{\textbf{k}} \equiv - 4 \tp \cos {k_x} \cos {k_y}$ and $\epsilon_{\textbf{k}} \equiv  - 2t(\cos{k_x} + \cos{k_y})$.

Suppose that the one particle eigenstates of the Hamiltonian are $|\gamma^\pm_{\textbf{k}\sigma}> = (P^{\pm}_{A \textbf{k}\sigma} \hat{c}_{A\textbf{k}  \sigma} \d + P^{\pm}_{B \textbf{k}\sigma} \hat{c}_{B\textbf{k}  \sigma} \d )|0>$ with eigenvalues $\tilde{\epsilon}^{\pm}_{\textbf{k}\sigma}$, which are clearly the {\em effective} one-particle band energies. Then the eigenvalue equation is
\be
\begin{pmatrix}
	A_{\textbf{k}  \sigma} & \epsilon_{\textbf{k}} \\
	\epsilon_{\textbf{k}} & B_{\textbf{k}  \sigma}\\
\end{pmatrix}
\begin{pmatrix}
	P^{\pm}_{A \textbf{k}\sigma}\\
	P^{\pm}_{B \textbf{k} \sigma}\\
\end{pmatrix}
= \tilde{\epsilon}^{\pm}_{\textbf{k}\sigma}
\begin{pmatrix}
	P^{\pm}_{A \textbf{k}\sigma}\\
	P^{\pm}_{B\textbf{k} \sigma}\\
\end{pmatrix}
\en
The band energies at wave-vector $\textbf{k}$ and for spin $\sigma$ are therefore given by
\be
\begin{split}
	&\tilde{\epsilon}^\pm_{\textbf{k}\sigma} = \frac{1}{2}\left( A_{\textbf{k}\sigma} + B_{\textbf{k}\sigma}\pm \sqrt{(A_{\textbf{k}\sigma} - B_{\textbf{k}\sigma})^2 + 4 (\epsilon_{\textbf{k}})^2} \right)\\
	& P^{\pm 2}_{B\textbf{k}\sigma} = 1/(1+L^{\pm 2}_{\textbf{k} \sigma})\\
	& P^{\pm 2}_{A\textbf{k}\sigma} = L^{\pm 2}_{\textbf{k} \sigma}/(1+L^{\pm 2}_{\textbf{k} \sigma})
\end{split}
\en
where $L^\pm_{\textbf{k} \sigma}  \equiv - \epsilon_{\textbf{k}} / [A_{\textbf{k}\sigma} - \tilde{\epsilon}^\pm_{\textbf{k}\sigma}] $. The particle occupancy at site $\alpha$ for the state ($|\gamma^\pm_{\textbf{k}\sigma}>$) is given by  $n^\pm_{\alpha \textbf{k} \sigma} = <\gamma^\pm_{\textbf{k}\sigma}|\hat{c}_{\alpha \textbf{k}  \sigma} \d \hat{c}_{\alpha \textbf{k}  \sigma} |\gamma^\pm_{\textbf{k}\sigma}> = P^{\pm 2}_{\alpha \textbf{k}\sigma}$.

In the ground state, only the single particle states with negative energies are occupied, Hence we get the following self consistency relations for the four occupancies at site $\alpha$ and for spin $\sigma$:
\be
	  n_{\alpha \sigma} = \sum_{\tilde{\epsilon}^\pm_{\textbf{k} \sigma} < 0} P^{\pm 2}_{\alpha \textbf{k}\sigma}
\en
For each set of parameter values $(U/t$, $\Delta/t$, and $\tp/t)$, we numerically solve the above four equations and the half filling constraint equation to determine the four (spin and site resolved) occupancies and the chemical potential $\mu$. The HF order parameter results we presented and discussed in section II are obtained from these occupancies using the relations
\begin{eqnarray}
% \nonumber % Remove numbering (before each equation)
  m_A &=& n_{A \uparrow} - n_{A \downarrow}, \; m_B = n_{B \uparrow} - n_{B \downarrow} \nonumber \\
  n_A &=& n_{A \uparrow} + n_{A \downarrow}, \; n_B = n_{B \uparrow} + n_{B \downarrow} \nonumber \\
  m_s &=& (m_A - m_B)/2, \; m_f = (m_A + m_B)/2  \nonumber \\
  \delta n &=& (n_A - n_B)/2 
\end{eqnarray}

We note here that to ensure the correctness of our HF calculations we have benchmarked them against 1) the $U_c$ Vs $\tp$ plot for $t-\tp$ Hubbard model in Vollhardt et al\cite{Vollhardt}. Furthermore, the analytical form of band gap in the non interacting $t-\tp$ IHM on a square lattice is $2\Delta - 4\tp$. The values of the gap calculated from the numerically calculated DOS from our code match well with this analytical form of the gap.

\section{The DMFT + CTQMC Method}
The formulation of the Dynamical Mean-field Theory (DMFT) for the IHM and the use of CTQMC as the impurity solver have been discussed in section II of our earlier work\cite{soumen} (corresponding to $\tp=0$) and in the references cited therein. In this appendix we confine ourselves to drawing attention to the specific differences that arise due to the presence of a non-zero $\tp$.
Even in the presence of $\tp$, the expression for the  $\textbf{k}$ and spin dependent $(2 \times 2)$ {\em matrix} Greens function for the IHM in terms of the local spin dependent self energies continues to be given by Eq. 2 of Ref. \cite{soumen}), i.e.,
\be
[\textbf{G}^\sigma (\textbf{k}, i\omega_n)]_{\alpha \beta} =
\begin{pmatrix}
	\zeta_{A \sigma} (\textbf{k}, i\omega_n) & \epsilon_{\textbf{k}} \\
	\epsilon_{\textbf{k}} & \zeta_{B \sigma} (\textbf{k}, i\omega_n)\\
\end{pmatrix}
^{-1} \; ,
\en
where the two-valued matrix indices $\alpha$ and $\beta$ correspond to the two sublattice indices (A, B). However, the diagonal entries on the right hand side are now $\textbf{k}$ dependent, due to the intra-sublattice hopping induced by $\tp$, where as the off-diagonal terms, due to the inter-sublattice hopping $t$, are given by $ - \epsilon_{\textbf{k}} = 2 t (\cos {k_x} + \cos {k_y})$ as before. For the diagonal terms we now have,
\be
\zeta_{\alpha \sigma} (\textbf{k}, i\omega_n) = \zeta_{\alpha \sigma} (i\omega_n) - \epsilon^{\prime}_{\textbf{k}}.
\en
Here, $\alpha = A, B$ is the sublattice index,
\be
\begin{split}
	\zeta_{A \sigma}(i\omega_n) &= i\omega_n + \mu - \Delta -\Sigma_{A,\sigma}(i\omega_n), \\
	\zeta_{B \sigma}(i\omega_n) &= i\omega_n + \mu + \Delta -\Sigma_{B,\sigma}(i\omega_n),
\end{split}
\en
in terms of the two local self energies $ \Sigma_{\alpha,\sigma}(i\omega_n)$ of the two separate self consistent impurity problems on sites A and B as before, and $\epsilon^{\prime}_{\textbf{k}} = - 4 \tp \cos {k_x} \cos {k_y}$. It is straightforward to verify that the diagonal components of the $\textbf{k}$ and spin dependent matrix Greens function are hence given by
\be
G_{\alpha \sigma}(\textbf{k}, i\omega_n) = \frac{\zeta_{\bar{\alpha} \sigma} (i\omega_n) - \epsilon'(\textbf{k})} {(\zeta_{A \sigma} (\textbf{k}, i\omega_n)- \epsilon^{\prime}_{\textbf{k}}) (\zeta_{B \sigma} (\textbf{k}, i\omega_n) - \epsilon^{\prime}_{\textbf{k}}) - \epsilon^2_{\textbf{k}}}
\en
where $\bar{\alpha} = B, A$ for $\alpha = A, B$.
The {\em local} Greens functions on an A or a B site are obtained by performing the $\textbf{k}$ sum,
\be
G_{\alpha \sigma}(i\omega_n) = \sum_{\textbf{k}} G_{\alpha \sigma}(\textbf{k}, i\omega_n)
\en
Unlike in the model with $\tp = 0$, the sum cannot be easily converted into an integral over a density of states as it involves two different energy dispersions.We carry it out numerically, using a $(2* L_g \times 2*L_g)$ grid over the square BZ. We used shifted k point gird where crystal momentum component $k_x$ and $k_y$ are defined as $\frac{(i-L_g)\pi}{L_g} + \frac{\pi}{2*L_g} $ where i runs from 0 to $(2*L_g-1)$. Because of the symmetry of Hamiltonian on the square lattice BZ, one does not need to sample the  whole BZ, sampling a 1/8 wedge of the BZ with appropriate weight will do the job. Momentum sums are carried out over these triangular wedges. (The triangular regime is bordered by $k_x=k_y$ where -$\pi < k_x <0$. k points on the $k_x=k_y$ line are ascribed weights of  4 and all other points in the triangle carry weights of 8.) We have verified the convergence of our calculations by checking that $L_g=1000$ yields the same results as the larger $L_g =2000$ and 4000, for several parameter sets in the HF case. All the reported data for both HF and CTQMC have been obtained using  $L_g = 1000.$

%\begin{lstlisting}
%	for(int i = 0; i<GRID; i++){ # k_x
%		for(int j = 0; j<(i+1); j++){ #k_y
%			
%			if(j!=i){weight=8;} 
%			else{weight=4;}
%			
%			\epsilon(\textbf{k}) = 2*t*(cos_[i] + cos_[j]);
%			\epsilon^{\prime}(\textbf{k}) = 2*dim*t2*cos_[i]*cos_[j];
%\end{lstlisting}

The hybridization functions on the A and B sites, which are the inputs to the CTQMC impurity solver, are then obtained as
\be
D_{\alpha,\sigma} (i \omega_n) = \zeta_{\alpha \sigma}(i\omega_n) - [G_{\alpha,\sigma}(i\omega_n)]^{-1}
\en
The CTQMC code (we use the CT-HYB package due to Kristjan Haule,et al \cite{ctqmchaule} that has been graciously made available in the public domain) then generates the local self energies on the A and B sites as functions of the Matsubara frequencies.
Starting from a trial self energy, eg., the Hartree-Fock self energies, the computation of the local greens functions, the Hybridization functions, and the local self energies are then iterated to convergence. Even if we start from a trial self energy that corresponds to a paramagnetic state, the CTQMC code we use are sophisticated enough to self-consistently generate non zero values for the local magnetizations, $m_A$ and $m_B$, as appropriate. The convergence criterion we use are that $n_A$ and $n_B$ are converged to an accuracy of 0.0002, and the updating of chemical potential is stopped once the two sublattice occupancy  $|(n_A+n_B)-2.0|$ reaches 0.001.
The momentum distribution function is calculable in a straightforward way using the converged self energies in the expression for the $\textbf{k}$, sublattice and spin resolved diagonal components of the matrix Greens function discussed above:
\be
n_{\alpha,\textbf{k},\sigma} =\frac{1}{\beta} \sum_{i\omega_n} G_{\alpha,\sigma}(\textbf{k}, i\omega_n) + \frac{1}{2}
\en

The Fermi surface is defined by the locus of points in the BZ where there is a jump in the momentum distribution function. These points can be located by determining where the gradient with respect to $\textbf{k}$ of $n_{\alpha,\textbf{k},\sigma}$ has the maximum magnitude. However, visually they are fairly easy to see in false color plots of the momentum distribution functions, as discussed in the text.

% K sum was done on GRID 2000*2000. CTQMC calculation was done with 10^7 Monte Carlo steps. for most os parameter in phase space it took around 150 self consistent DMFT cycle. For parameter near phase transition it took around 200 self consistent cycle.  

 From the $n_\textbf{k}$ data a transition from the AFHM phase to the AFI phase is inferred when $n_\textbf{k} $ differs from 0.5 by less than 0.02. The $U_c$ line corresponding to the  PM $\rightarrow$ FM continuous transition is inferred when $m_s $ and $m_f$ become more than 0.001. The BI to metal transition is inferred when $n_\textbf{k}$ differs from 0.5 by more than 0.02.


\begin{thebibliography}{99}
\bibitem{foot_note2}	
	 In this paper we restrict our attention to positive values of $\tp$. The physics for negative values of $\tp$ is obtainable simply via a particle-hole transformation.
\end{thebibliography}
\end{document}